\documentclass[aps,amsmath,amssymb, reprint, prl, preprintnumbers]{revtex4-1}
\pdfoutput=1

\usepackage{graphicx}
\usepackage{epstopdf}
\usepackage{amsmath, amssymb}

\usepackage{hyperref}
\usepackage{bbm,array,amsfonts,graphicx,wrapfig,float,mathtools,multirow}
\usepackage[dvipsnames]{xcolor}

\newcommand{\be}{\begin{equation}}
\newcommand{\ee}{\end{equation}}
\newcommand{\beq}{\begin{equation}}
\newcommand{\beql}[1]{\begin{equation}\label{#1}}
\newcommand{\eeq}{\end{equation}}
\newcommand{\ba}{\begin{array}}
\newcommand{\ea}{\end{array}}
\newcommand{\bea}{\begin{eqnarray}}
\newcommand{\beal}[1]{\begin{eqnarray}\label{#1}}
\newcommand{\eea}{\end{eqnarray}}
\newcommand{\ben}{\begin{enumerate}}
\newcommand{\een}{\end{enumerate}}
\newcommand{\bean}{\begin{eqnarray*}}
\newcommand{\eean}{\end{eqnarray*}}
\newcommand{\eref}[1]{(\ref{#1})}

\newcommand{\tref}[1]{Table~\ref{#1}}
\newcommand{\nn}{\nonumber}

\newcommand{\fref}[1]{Figure \ref{#1}}

\begin{document}

\title{Quiver-Invariant Dualities between Brane Tilings}

\preprint{UNIST-MTH-26-RS-01}
\preprint{CGP26002}

\author{Minsung Kho}
\email[\texttt{minsung@unist.ac.kr}]{}
\affiliation{
Department of Mathematical Sciences, Ulsan National Institute of Science and Technology,
50 UNIST-gil, Ulsan 44919, South Korea
}

\author{Seong-Jin Lee}
\email[\texttt{seongjinlee@ibs.re.kr}]{}
\affiliation{
Center for Geometry and Physics, Institute for Basic Science (IBS),
Pohang 37673, South Korea
}

\author{Rak-Kyeong Seong}
\email[\texttt{seong@unist.ac.kr}]{}
\affiliation{
Department of Mathematical Sciences, and Department of Physics, Ulsan National Institute of Science and Technology,
50 UNIST-gil, Ulsan 44919, South Korea
}

\begin{abstract}
We study pairs of
$4d$ $\mathcal{N}=1$ supersymmetric gauge theories
that share the same vacuum moduli space
and the same chiral field content, encoded by a common quiver, 
but differ in their superpotentials.
These theories arise
as worldvolume theories on a D3-brane probing a toric Calabi-Yau 3-fold and
admit a description in terms of bipartite graphs on a 2-torus, known as brane tilings.
Using an explicit example, we show that the correspondence
is realized by a single `tilting' mutation along the diagonals of hexagonal faces in the brane tiling,
which is equivalent to a specific sequence of Seiberg dualities performed at distinct gauge nodes in the quiver.
\end{abstract} 

\maketitle

\section{Introduction}

Worldvolume theories on D3-branes probing toric Calabi-Yau 3-folds
form a large family of $4d$ $\mathcal{N}=1$ supersymmetric quiver gauge theories, 
which have been studied 
extensively from various perspectives.
A notable feature of this family 
is that
both  their Lagrangians and the underlying Type IIB brane configurations can be encoded
by a bipartite graph on a 2-torus $T^2$ known as a brane tiling \cite{Franco:2005rj, Hanany:2005ve, Franco:2005sm} or a dimer model 
\cite{2003math.....10326K,kasteleyn1967graph}.
Another key feature
is that these $4d$ $\mathcal{N}=1$
theories exhibit Seiberg duality \cite{Seiberg:1994pq}, which in the context of brane tilings is also referred to as toric duality \cite{Feng:2000mi, Feng:2001bn, Feng:2001xr, Feng:2002zw}.

\begin{figure}[ht!!]
\begin{center}
\resizebox{0.75\hsize}{!}{
\includegraphics[height=5cm]{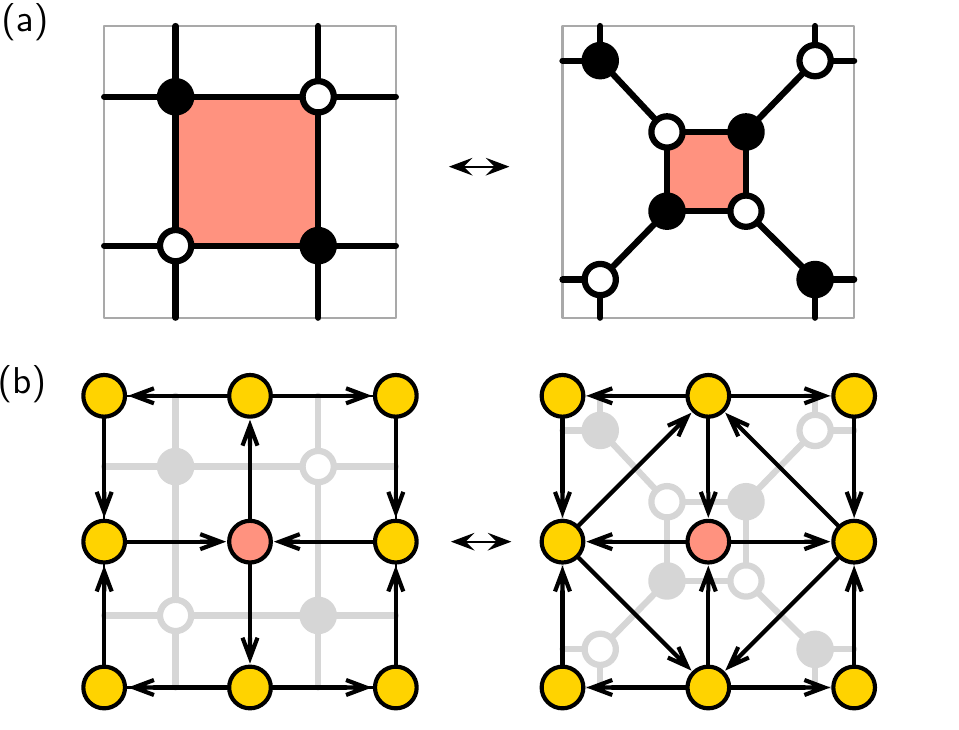} 
}
\caption{
Seiberg duality 
acts on (a) quadrilateral faces (red) of a brane tiling and 
(b) corresponding gauge nodes (red) in the periodic quiver of the associated $4d$ $\mathcal{N}=1$ theory. 
The local mutation on the brane tiling and the periodic quiver is also referred to as a spider move or urban renewal \cite{2011arXiv1107.5588G, CIUCU199834, 1999math......3025K}.
\label{fig_00}}
 \end{center}
 \end{figure}

Two $4d$ $\mathcal{N}=1$ theories 
are Seiberg dual if they flow to the same IR fixed point under a renormalization group flow.
In terms of brane tilings, 
Seiberg duality is realized as a
local mutation of the bipartite graph on $T^2$, commonly known as a
spider move or urban renewal \cite{2011arXiv1107.5588G, CIUCU199834, 1999math......3025K}, as illustrated in \fref{fig_00}.
This mutation acts on quadrilateral faces of the brane tiling, where faces correspond to $U(N)$ gauge groups in the corresponding $4d$ $\mathcal{N}=1$ theory. 
A key feature is that Seiberg duality leaves the vacuum moduli space -- the space of solutions to the $F$- and $D$-terms of the 
$4d$ $\mathcal{N}=1$ theory, also known as the mesonic moduli space \cite{Butti:2007jv, Forcella:2008bb, Forcella:2008eh} -- invariant. 
While related brane tiling mutations have been studied \cite{Bianchi:2014qma, 2022SIGMA..18..030H, Franco:2023flw, Cremonesi:2023psg, Arias-Tamargo:2024fjt, Franco:2023mkw,CarrenoBolla:2024fxy, Kho:2025fmp}, for instance in the context of specular duality \cite{Hanany:2012vc},
none were found to preserve the mesonic moduli space in the same way as Seiberg duality.

\begin{figure}[htt!!]
\begin{center}
\resizebox{1\hsize}{!}{
\includegraphics[height=5cm]{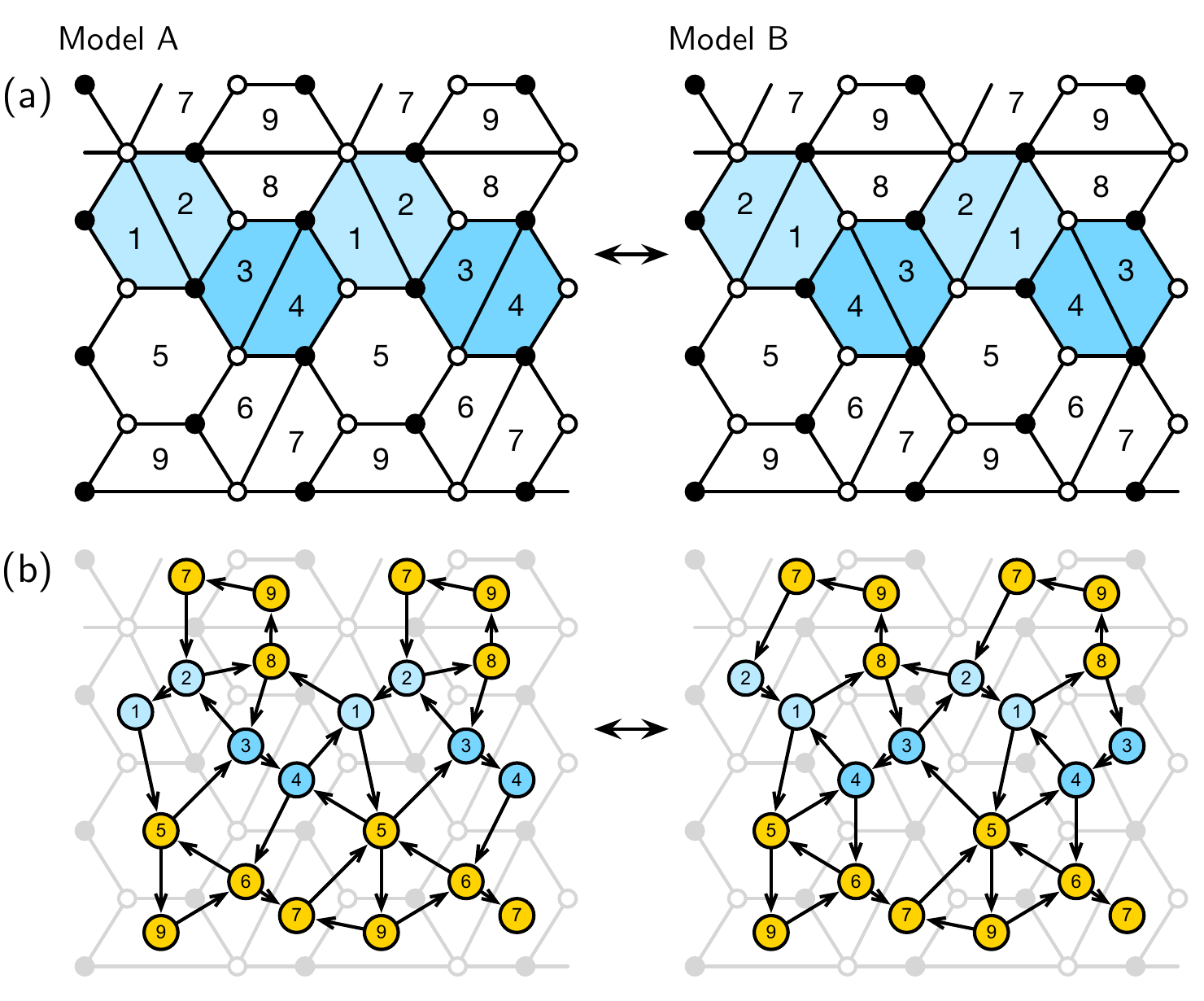} 
}
\caption{
(a) The local `tilting' mutation acts along the diagonals of hexagonal faces (blue) in the brane tiling. 
(b) The mutation modifies the periodic quiver on the 2-torus $T^2$ and hence the superpotential, 
while leaving the underlying quiver and the mesonic moduli space of the associated $4d$ $\mathcal{N}=1$ theories invariant. 
The `tilting’ mutation realizes a particular sequence of Seiberg dualities.
\label{f_fig04}}
 \end{center}
 \end{figure}

The local mutation of the quiver associated with Seiberg duality
modifies the directed graph, so that the two quivers related by this mutation are not equivalent.  
In this work, we report a family of 
dualities between 
$4d$ $\mathcal{N}=1$ theories given by brane tilings
that share the same quiver and chiral field content,
but differ in their superpotentials.
The theories related by this duality have the same mesonic moduli space.
This suggests that
they are not related by a single Seiberg duality -- which would typically change the shape of the quiver -- 
but rather by a sequence of Seiberg dualities that, remarkably, leaves the quiver, as a directed graph, invariant.

We show that this new correspondence is realized as
a single `tilting' mutation along diagonals of hexagonal faces in the associated brane tilings, as illustrated in \fref{f_fig04}.
By treating the bipartite graph of the brane tiling purely as a statistical mechanics model on a lattice \cite{1997AIHPB..33..591K},
this mutation preserves the nearest-neighbor adjacency of faces, reflecting the invariance of the quiver.
Moreover, this local `tilting' move is equivalent to a specific sequence of
spider moves, corresponding to multiple
Seiberg dualities acting on different $U(N)$ gauge nodes in the quiver. 
We emphasize here that the `tilting' mutation leaves the quiver invariant, including the labels on gauge nodes, 
whereas the sequence of Seiberg dualities only leaves the shape of the quiver invariant, requiring possibly an additional relabelling of nodes to be fully equivalent to the `tilting' mutation.

\section{$4d$ $\mathcal{N}=1$ Theories and Brane Tilings}

\begin{figure}[ht!!]
\begin{center}
\resizebox{0.8\hsize}{!}{
\includegraphics[height=5cm]{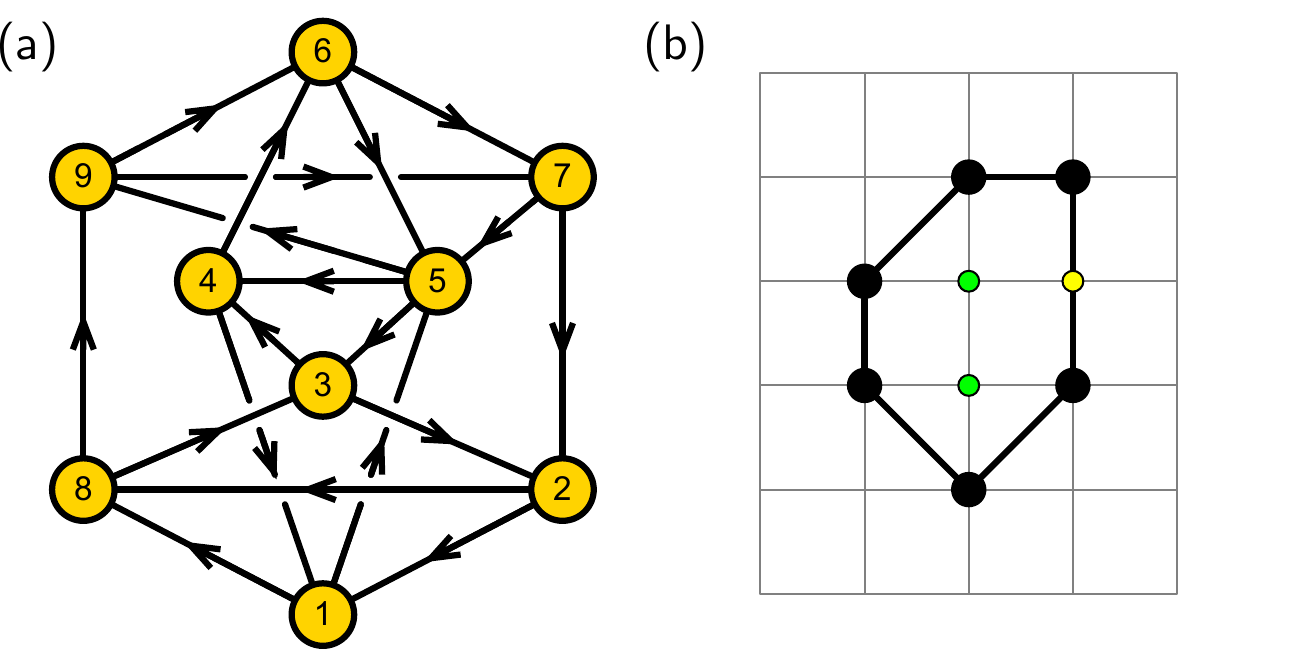} 
}
\caption{
(a) Quiver
and (b) toric diagram of the toric Calabi-Yau 3-fold
shared by Models A and B. 
\label{fig_01}}
 \end{center}
 \end{figure}

We present an explicit example of the new correspondence
by presenting two $4d$ $\mathcal{N}=1$ theories described by brane tilings.
The shared quiver is shown in \fref{fig_01}, together with the toric diagram \cite{fulton1993introduction, cox2011graduate} of the associated toric Calabi-Yau 3-fold.
For an abelian $4d$ $\mathcal{N}=1$ theory with $U(1)$ gauge groups,
the mesonic moduli space $\mathcal{M}^{mes}$ is precisely the probed toric Calabi-Yau 3-fold itself, 
whereas for $U(N)$ gauge groups the mesonic moduli space becomes an $N$-th symmetric product of $\mathcal{M}^{mes}$.
The mesonic moduli space \cite{Butti:2007jv, Forcella:2008bb, Forcella:2008eh} for an abelian $4d$ $\mathcal{N}=1$ theory described by a brane tiling is defined as, 
\beal{es01a01}
\mathcal{M}^{mes} = \text{Spec} (\mathbb{C}[X_{ij}] / \mathcal{I}_{Irr}) // U(1)^{G-1} 
~,~
\eea
where the coordinate ring $\mathbb{C}[X_{ij}]$, written in terms of bifundamental chiral fields $X_{ij}$,
is subject to the irreducible principal component $\mathcal{I}_{Irr}$ of the binomial ideal formed by the $F$-term relations.
The overall quotient by the independent $U(1)^{G-1}$ gauge symmetries ensures that $\mathcal{M}^{mes}$ captures gauge invariant operators, where $G$ is the number of gauge groups with an overall decoupled $U(1)$. 

In terms of GLSM fields $p_{a}$ \cite{Witten:1993yc} corresponding to vertices of the toric diagram, 
the mesonic moduli space $\mathcal{M}^{mes}$ can be expressed as a symplectic quotient of the following form, 
\beal{es01a05}
\mathcal{M}^{mes} = \text{Spec} (\mathbb{C}[p_a] // Q_D // Q_F) ~,~
\eea
where the coordinate ring $\mathbb{C}[p_a]$ 
is subject to $D$-terms encoded in a $U(1)$ charge matrix $Q_D$
and $F$-terms encoded in a $U(1)$ charge matrix $Q_F$. 
The chiral fields $X_{ij}$
can be expressed in terms of the GLSM fields $p_a$, which are combinatorially encoded in
the brane tiling as perfect matchings of the bipartite graph.
The $P$-matrix encodes the relation between GLSM fields and chiral fields and is obtained 
via the forward algorithm \cite{Feng:2000mi} from the $F$-terms. Combined with the $U(1)$ gauge charges, 
the $P$-matrix allows one to obtain the charge matrices $Q_D$ and $Q_F$ following the forward algorithm.
The kernel of the combined charge matrix $Q_t=(Q_D, Q_F)$ is, 
\beal{es01a06}
G_t = \text{ker}(Q_t) ~,~
\eea
and its columns give the coordinates of the toric diagram vertices associated with the GLSM fields.

We also present the Hilbert series of $\mathcal{M}^{mes}$ \cite{Benvenuti:2006qr, Hanany:2006uc, Feng:2007ur, Butti:2007jv},
which is the generating function for mesonic gauge invariant operators of the brane tiling.
It is defined as, 
\beal{es01a10}
&&
g(t_a; \mathcal{M}^{mes})
= 
\nn\\
&&
\hspace{0.5cm}
\prod_{m=1}^{c-3} \oint_{|z_m|=1} \frac{d z_m}{2\pi i z_m} \prod_{a=1}^{c}
\frac{1}{1-t_a \prod_{n=1}^{c-3}z_n^{(Q_t)_{na}}}
~,~
\eea
where $c$ corresponds to the number of GLSM fields $p_a$ and $t_a$ is the fugacity counting degrees in $p_a$.
The Hilbert series allows us to
calculate also the set of generators for $\mathcal{M}^{mes}$ using plethystics \cite{Benvenuti:2006qr, Feng:2007ur}. 
In our example, 
we find that the two $4d$ $\mathcal{N}=1$ theories related by the new correspondence
share the same set of generators, since $\mathcal{M}^{mes}$ is the same. 
Even though the 
chiral field content and hence the quiver are identical, 
the generators for $\mathcal{M}^{mes}$ are
realized differently in terms of the chiral fields in the two theories.

We also observe that the global symmetry is the same for the two brane tilings. 
Accordingly, the generators of the invariant mesonic moduli space $\mathcal{M}^{mes}$ carry
the same charges under the global symmetry, including identical superconformal $U(1)_R$ charges. 
By calculating the $U(1)_R$ charges using $a$-maximization \cite{Intriligator:2003jj, Butti:2005vn, Butti:2005ps}, 
we show that the generators of $\mathcal{M}^{mes}$ have the same $U(1)_R$ charges, 
while the chiral fields themselves do not necessarily carry the same $U(1)_R$ charges in the two $4d$ $\mathcal{N}=1$ theories.

\section{A New Correspondence}

\subsection{Model A}

\begin{figure}[H]
\begin{center}
\resizebox{0.85\hsize}{!}{
\includegraphics[height=5cm]{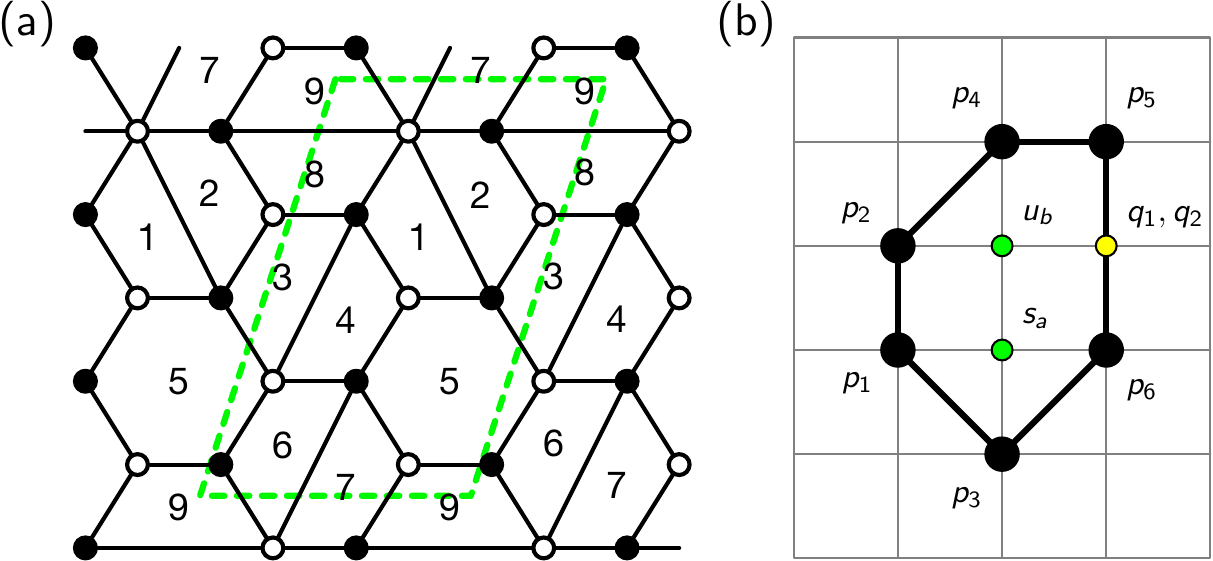} 
}
\caption{
(a) The brane tiling and (b) the labelled toric diagram
for Model A, where
vertices in the toric diagram correspond to GLSM fields $p_a$.
The dotted quadrilateral in the brane tiling indicates the fundamental domain on the 2-torus $T^2$.
\label{fig_02}}
 \end{center}
 \end{figure}

\begin{table}[ht!!]
\centering
\begin{tabular}{|c|c|c|c|c|}
\hline
\; & $U(1)_{f_1}$ & $U(1)_{f_2}$ & $U(1)_R$ & fugacity \\
\hline \hline
$p_1$ &  1	&  0	& $R_1\simeq 0.273550 $ & $t_1$ \\
$p_2$ &  1 	&  0	& $R_2\simeq 0.320461$ & $t_2$ \\
$p_3$ & -1        &  1 	& $R_3\simeq 0.386171$ & $t_3$ \\
$p_4$ & -1        &  1 	& $R_4\simeq 0.253224$ & $t_4$ \\
$p_5$ &  0        &  -1	& $R_5\simeq 0.404904$ & $t_5$ \\
$p_6$ &  0        &  -1	& $R_6\simeq 0.361689$ & $t_6$ \\
\hline
\end{tabular}
\caption{The GLSM fields $p_a$ corresponding to extremal points of the toric diagram with their mesonic flavor charges
and $U(1)_R$ charges $R_a$, where $\sum_a R_a = 2$.
These $U(1)_R$ charges correspond to the minimum volume of divisors in the associated toric Calabi-Yau 3-fold.
The $U(1)_R$ charges on GLSM fields $p_a$ are the same for Models A and B.
 \label{tab_01}}
\end{table}

\begin{table}[ht!!]
\centering
\begin{tabular}{|c|c|c|l|}
\hline
\; & $U(1)_{f_1}$ & $U(1)_{f_2}$ & $U(1)_R$  \\
\hline \hline
$X_{15}$ &  -1        &  0	& $R_3+R_6\simeq 0.74786$  \\
$X_{18}$ &  +1        &  0	& $R_2\simeq 0.320461$ \\
$X_{21}$ &  +1        &  0	& $R_1\simeq 0.273550$  \\
$X_{28}$ &  0        &  -2	& $R_5+R_6\simeq 0.766593$  \\
$X_{32}$ &  0        &  +1	& $R_2+R_4\simeq 0.573685$  \\
$X_{34}$ &  0        &  -1	& $R_6\simeq 0.361689$  \\
$X_{41}$ &  -1        &  0	& $R_4+R_5\simeq 0.658128$  \\
$X_{53}$ &  0        &  -1	& $R_5\simeq 0.404904$  \\
$X_{54}$ &  +2        &  0	& $R_1+R_2\simeq 0.594011$  \\
$X_{83}$ &  0        &  +1	& $R_1+R_3\simeq 0.659722$  \\
\hline
$X_{46}$ &  -1        &  +1	& $R_3\simeq 0.386171$  \\
$X_{59}$ &  -1        &  0	& $R_3+R_6\simeq 0.747860$  \\
$X_{65}$ &  +1        &  +1	& $R_1+R_2+R_4\simeq 0.847235$ \\
$X_{67}$ &  0        &  -1	& $R_6\simeq 0.361689$  \\
$X_{72}$ &  -1        &  +1	& $R_3\simeq 0.386171$  \\
$X_{75}$ &  -1        &  0	& $R_4+R_5\simeq 0.658128$  \\
$X_{89}$ &  -1        &  +1	& $R_4\simeq 0.253224$  \\
$X_{96}$ &  0 	     &  -1	& $R_5\simeq 0.404904$  \\
$X_{97}$ &  +2	     &  0	& $R_1+R_2\simeq 0.594011 $  \\
\hline
\end{tabular}
\caption{
The charges under the global symmetry on chiral fields in Model A, including the $U(1)_R$ charges that are expressed in 
terms of $U(1)_R$ charges $R_a$ corresponding to the extremal GLSM fields $p_a$.
Note that the global symmetry charges of the chiral fields $X_{46}, \dots, X_{97}$ are identical to those in Model B, whereas the charges of the remaining chiral fields -- those affected by the `tilting’ mutation of the brane tiling -- differ between the two models (up to a re-ordering of charges).
 \label{tab_02}}
\end{table}

The quiver for Model A
is shown in \fref{fig_01}
and the superpotential is given by,
\beal{es01a01}
&&
W^{A}
=
X_{97} X_{75} X_{59}+X_{54} X_{41} X_{15}+X_{28} X_{83} X_{32}
\nn
\\
&&
\hspace{0.25cm}
+X_{53}X_{34}X_{46}X_{65}+X_{96}X_{67}X_{72}X_{21}X_{18}X_{89}
\nn
\\
&&
\hspace{0.25cm}
-X_{96} X_{65} X_{59}- X_{97} X_{72} X_{28} X_{89}- X_{75} X_{54} X_{46} X_{67}
\nn
\\
&&
\hspace{0.25cm}
- X_{53} X_{32} X_{21} X_{15}-X_{41}X_{18}X_{83}X_{34}
~.~
\eea
The corresponding brane tiling is shown in \fref{fig_02}.
Using the forward algorithm, we obtain the
perfect matching matrix $P^{A}$ for Model A,
\beal{es01a02}
P^{A}=
\resizebox{0.37\textwidth}{!}{$
\left(
\ba{c|cccccc|cc|cccccccccc|cccccccccccc}
\; &  p_1 & p_2 & p_3 & p_4 & p_5 & p_6 & q_1 & q_2 & s_1 & s_2 & s_3 & s_4 & s_5 & s_6 & s_7 & s_8 & s_9 & s_{10} & u_1 & u_2 & u_3 & u_4 & u_5 & u_6 & u_7 & u_8 & u_9 & u_{10} & u_{11} & u_{12}  \\
\hline
 X_{15} &   0 & 0 & 1 & 0 & 0 & 1 & 0 & 1 & 1 & 1 & 0 & 0 & 0 & 0 & 0 & 1 & 1 & 1 & 0 & 0 & 0 & 0 & 1 & 1 & 0 & 0 & 0 & 0 & 0 & 0 \\
 X_{18} &   0 & 1 & 0 & 0 & 0 & 0 & 0 & 0 & 0 & 1 & 0 & 0 & 0 & 0 & 0 & 0 & 0 & 0 & 0 & 0 & 0 & 1 & 0 & 1 & 0 & 0 & 0 & 0 & 0 & 0 \\
 X_{21} &   1 & 0 & 0 & 0 & 0 & 0 & 0 & 0 & 0 & 0 & 1 & 1 & 0 & 0 & 0 & 0 & 0 & 0 & 0 & 0 & 0 & 0 & 0 & 0 & 1 & 0 & 0 & 0 & 0 & 0 \\
 X_{28} &   0 & 0 & 0 & 0 & 1 & 1 & 1 & 1 & 0 & 1 & 1 & 1 & 0 & 0 & 0 & 0 & 0 & 0 & 0 & 0 & 0 & 1 & 0 & 1 & 1 & 0 & 0 & 0 & 0 & 0 \\
 X_{34} &   0 & 0 & 0 & 0 & 0 & 1 & 0 & 1 & 0 & 0 & 0 & 1 & 0 & 1 & 0 & 0 & 0 & 0 & 0 & 1 & 0 & 0 & 0 & 0 & 0 & 0 & 0 & 0 & 1 & 0 \\
 X_{32} &   0 & 1 & 0 & 1 & 0 & 0 & 0 & 0 & 0 & 0 & 0 & 0 & 1 & 1 & 0 & 0 & 0 & 0 & 1 & 1 & 0 & 0 & 0 & 0 & 0 & 1 & 1 & 1 & 1 & 0 \\
 X_{41} &   0 & 0 & 0 & 1 & 1 & 0 & 1 & 0 & 0 & 0 & 1 & 0 & 1 & 0 & 0 & 0 & 0 & 0 & 1 & 0 & 0 & 0 & 0 & 0 & 1 & 1 & 1 & 1 & 0 & 0 \\
 X_{53} &   0 & 0 & 0 & 0 & 1 & 0 & 1 & 0 & 0 & 0 & 0 & 0 & 0 & 0 & 1 & 0 & 0 & 0 & 0 & 0 & 1 & 1 & 0 & 0 & 0 & 0 & 0 & 0 & 0 & 1 \\
 X_{54} &   1 & 1 & 0 & 0 & 0 & 0 & 0 & 0 & 0 & 0 & 0 & 1 & 0 & 1 & 1 & 0 & 0 & 0 & 0 & 1 & 1 & 1 & 0 & 0 & 0 & 0 & 0 & 0 & 1 & 1 \\
 X_{83} &   1 & 0 & 1 & 0 & 0 & 0 & 0 & 0 & 1 & 0 & 0 & 0 & 0 & 0 & 1 & 1 & 1 & 1 & 0 & 0 & 1 & 0 & 1 & 0 & 0 & 0 & 0 & 0 & 0 & 1 \\
 X_{46} &   0 & 0 & 1 & 0 & 0 & 0 & 0 & 0 & 1 & 1 & 1 & 0 & 1 & 0 & 0 & 0 & 0 & 1 & 1 & 0 & 0 & 0 & 0 & 0 & 0 & 0 & 0 & 1 & 0 & 0 \\
 X_{59} &   0 & 0 & 1 & 0 & 0 & 1 & 1 & 0 & 1 & 1 & 1 & 1 & 1 & 1 & 1 & 0 & 0 & 0 & 1 & 1 & 1 & 1 & 0 & 0 & 0 & 0 & 0 & 0 & 0 & 0 \\
 X_{65} &   1 & 1 & 0 & 1 & 0 & 0 & 0 & 0 & 0 & 0 & 0 & 0 & 0 & 0 & 0 & 1 & 1 & 0 & 0 & 0 & 0 & 0 & 1 & 1 & 1 & 1 & 1 & 0 & 0 & 0 \\
 X_{67} &   0 & 0 & 0 & 0 & 0 & 1 & 1 & 0 & 0 & 0 & 0 & 0 & 0 & 0 & 0 & 0 & 1 & 0 & 0 & 0 & 0 & 0 & 0 & 0 & 0 & 0 & 1 & 0 & 0 & 0 \\
 X_{72} &   0 & 0 & 1 & 0 & 0 & 0 & 0 & 0 & 0 & 0 & 0 & 0 & 1 & 1 & 1 & 1 & 0 & 0 & 0 & 0 & 0 & 0 & 0 & 0 & 0 & 1 & 0 & 0 & 0 & 0 \\
 X_{75} &   0 & 0 & 0 & 1 & 1 & 0 & 0 & 1 & 0 & 0 & 0 & 0 & 0 & 0 & 0 & 1 & 0 & 0 & 0 & 0 & 0 & 0 & 1 & 1 & 1 & 1 & 0 & 0 & 0 & 0 \\
 X_{89} &   0 & 0 & 0 & 1 & 0 & 0 & 0 & 0 & 1 & 0 & 0 & 0 & 0 & 0 & 0 & 0 & 0 & 0 & 1 & 1 & 1 & 0 & 1 & 0 & 0 & 0 & 0 & 0 & 0 & 0 \\
 X_{96} &   0 & 0 & 0 & 0 & 1 & 0 & 0 & 1 & 0 & 0 & 0 & 0 & 0 & 0 & 0 & 0 & 0 & 1 & 0 & 0 & 0 & 0 & 0 & 0 & 0 & 0 & 0 & 1 & 1 & 1 \\
 X_{97} &   1 & 1 & 0 & 0 & 0 & 0 & 0 & 0 & 0 & 0 & 0 & 0 & 0 & 0 & 0 & 0 & 1 & 1 & 0 & 0 & 0 & 0 & 0 & 0 & 0 & 0 & 1 & 1 & 1 & 1 
\ea
\right)
$}
~.~
\eea
The corresponding F-term charge matrix $Q_F^{A}$ is given by,
\beal{es01a03}
Q_F^{A}=
\resizebox{0.37\textwidth}{!}{$
\left(
\ba{cccccc|cc|cccccccccc|cccccccccccc}
  p_1 & p_2 & p_3 & p_4 & p_5 & p_6 & q_1 & q_2 & s_1 & s_2 & s_3 & s_4 & s_5 & s_6 & s_7 & s_8 & s_9 & s_{10} & u_1 & u_2 & u_3 & u_4 & u_5 & u_6 & u_7 & u_8 & u_9 & u_{10} & u_{11} & u_{12}  \\
\hline
 1 & 0 & 0 & 0 & 0 & 0 & 0 & 1 & 0 & 0 & 0 & 0 & 0 & 0 & 0 & 0 & 0 & -1 & 0 & 0 & 0 & 0 & 0 & 0 & -1 & 0 & 0 & 1 & -1 & 0 \\
 0 & 1 & 0 & 0 & 0 & 0 & 0 & 1 & 0 & 0 & 0 & 0 & 0 & 0 & 0 & 0 & 0 & 0 & 0 & 0 & 0 & 0 & 0 & -1 & 0 & 0 & 0 & 0 & -1 & 0 \\
 0 & 0 & 1 & 0 & 0 & 0 & 0 & 0 & 0 & 0 & 0 & 0 & 0 & 0 & 0 & 0 & 0 & -2 & 0 & 0 & 0 & -1 & 0 & 1 & 0 & -1 & 0 & 1 & 0 & 1 \\
 0 & 0 & 0 & 1 & 0 & 0 & 0 & 0 & 0 & 0 & 0 & 0 & 0 & 0 & 0 & 0 & 0 & 1 & 0 & 0 & 0 & 0 & -1 & 0 & 0 & 0 & 0 & -1 & 0 & 0 \\
 0 & 0 & 0 & 0 & 1 & 0 & 0 & -1 & 0 & 0 & 0 & 0 & 0 & 0 & 0 & 0 & 0 & 1 & 0 & 0 & 0 & 0 & 0 & 0 & 0 & 0 & 0 & -1 & 1 & -1 \\
 0 & 0 & 0 & 0 & 0 & 1 & 0 & -1 & 0 & 0 & 0 & 0 & 0 & 0 & 0 & 0 & 0 & -1 & 0 & 0 & 0 & -1 & 0 & 1 & 0 & 0 & -1 & 1 & 0 & 1 \\
 0 & 0 & 0 & 0 & 0 & 0 & 1 & -1 & 0 & 0 & 0 & 0 & 0 & 0 & 0 & 0 & 0 & 0 & 0 & 0 & 0 & -1 & 0 & 1 & 0 & 0 & -1 & 0 & 1 & 0 \\
 0 & 0 & 0 & 0 & 0 & 0 & 0 & 0 & 1 & 0 & 0 & 0 & 0 & 0 & 0 & 0 & 0 & -1 & 0 & 0 & 0 & -1 & -1 & 1 & 0 & 0 & 0 & 0 & 0 & 1 \\
 0 & 0 & 0 & 0 & 0 & 0 & 0 & 0 & 0 & 1 & 0 & 0 & 0 & 0 & 0 & 0 & 0 & -1 & 0 & 0 & 0 & -1 & 0 & 0 & 0 & 0 & 0 & 0 & 0 & 1 \\
 0 & 0 & 0 & 0 & 0 & 0 & 0 & 0 & 0 & 0 & 1 & 0 & 0 & 0 & 0 & 0 & 0 & -1 & 0 & 0 & 0 & -1 & 0 & 1 & -1 & 0 & 0 & 0 & 0 & 1 \\
 0 & 0 & 0 & 0 & 0 & 0 & 0 & 0 & 0 & 0 & 0 & 1 & 0 & 0 & 0 & 0 & 0 & -1 & 0 & 0 & 0 & -1 & 0 & 1 & -1 & 0 & 0 & 1 & -1 & 1 \\
 0 & 0 & 0 & 0 & 0 & 0 & 0 & 0 & 0 & 0 & 0 & 0 & 1 & 0 & 0 & 0 & 0 & -1 & 0 & 0 & 0 & -1 & 0 & 1 & 0 & -1 & 0 & 0 & 0 & 1 \\
 0 & 0 & 0 & 0 & 0 & 0 & 0 & 0 & 0 & 0 & 0 & 0 & 0 & 1 & 0 & 0 & 0 & -1 & 0 & 0 & 0 & -1 & 0 & 1 & 0 & -1 & 0 & 1 & -1 & 1 \\
 0 & 0 & 0 & 0 & 0 & 0 & 0 & 0 & 0 & 0 & 0 & 0 & 0 & 0 & 1 & 0 & 0 & -1 & 0 & 0 & 0 & -1 & 0 & 1 & 0 & -1 & 0 & 1 & 0 & 0 \\
 0 & 0 & 0 & 0 & 0 & 0 & 0 & 0 & 0 & 0 & 0 & 0 & 0 & 0 & 0 & 1 & 0 & -1 & 0 & 0 & 0 & 0 & 0 & 0 & 0 & -1 & 0 & 1 & 0 & 0 \\
 0 & 0 & 0 & 0 & 0 & 0 & 0 & 0 & 0 & 0 & 0 & 0 & 0 & 0 & 0 & 0 & 1 & -1 & 0 & 0 & 0 & 0 & 0 & 0 & 0 & 0 & -1 & 1 & 0 & 0 \\
 0 & 0 & 0 & 0 & 0 & 0 & 0 & 0 & 0 & 0 & 0 & 0 & 0 & 0 & 0 & 0 & 0 & 0 & 1 & 0 & 0 & -1 & -1 & 1 & 0 & 0 & 0 & -1 & 0 & 1 \\
 0 & 0 & 0 & 0 & 0 & 0 & 0 & 0 & 0 & 0 & 0 & 0 & 0 & 0 & 0 & 0 & 0 & 0 & 0 & 1 & 0 & -1 & -1 & 1 & 0 & 0 & 0 & 0 & -1 & 1 \\
 0 & 0 & 0 & 0 & 0 & 0 & 0 & 0 & 0 & 0 & 0 & 0 & 0 & 0 & 0 & 0 & 0 & 0 & 0 & 0 & 1 & -1 & -1 & 1 & 0 & 0 & 0 & 0 & 0 & 0 \\
\ea
\right)
$}
~,~
\eea
and the associated 
D-term charge matrix $Q_D^{A}$ takes the form,
\beal{es01a04}
Q_D^{A}=
\resizebox{0.37\textwidth}{!}{$
\left(
\ba{cccccc|cc|cccccccccc|cccccccccccc}
  p_1 & p_2 & p_3 & p_4 & p_5 & p_6 & q_1 & q_2 & s_1 & s_2 & s_3 & s_4 & s_5 & s_6 & s_7 & s_8 & s_9 & s_{10} & u_1 & u_2 & u_3 & u_4 & u_5 & u_6 & u_7 & u_8 & u_9 & u_{10} & u_{11} & u_{12}  \\
\hline
 0 & 0 & 1 & 0 & 0 & 0 & 0 & 0 & 0 & 0 & 0 & 0 & 0 & 0 & 0 & 0 & -2 & 0 & 0 & 0 & 0 & 0 & 1 & 0 & 0 & 0 & 0 & 0 & 0 & 0 \\
 0 & 0 & 0 & 0 & 0 & 1 & 0 & -1 & 0 & 0 & 0 & 0 & 0 & 0 & 0 & 0 & -1 & 0 & 0 & 0 & 0 & 0 & 1 & 0 & 0 & 0 & 0 & 0 & 0 & 0 \\
 0 & 0 & 0 & 0 & 0 & 0 & 1 & -1 & 0 & 0 & 0 & 0 & 0 & 0 & 0 & 0 & 0 & 0 & 0 & 0 & 0 & 0 & 0 & 0 & 0 & 0 & 0 & 0 & 0 & 0 \\
 0 & 0 & 0 & 0 & 0 & 0 & 0 & 0 & 1 & 0 & 0 & 0 & 0 & 0 & 0 & 0 & -1 & 0 & 0 & 0 & 0 & 0 & 0 & 0 & 0 & 0 & 0 & 0 & 0 & 0 \\
 0 & 0 & 0 & 0 & 0 & 0 & 0 & 0 & 0 & 1 & 0 & 0 & 0 & 0 & 0 & 0 & -1 & 0 & 0 & 0 & 0 & 0 & 0 & 0 & 0 & 0 & 0 & 0 & 0 & 0 \\
 0 & 0 & 0 & 0 & 0 & 0 & 0 & 0 & 0 & 0 & 1 & 0 & 0 & 0 & 0 & 0 & -1 & 0 & 0 & 0 & 0 & 0 & 0 & 0 & 0 & 0 & 0 & 0 & 0 & 0 \\
 0 & 0 & 0 & 0 & 0 & 0 & 0 & 0 & 0 & 0 & 0 & 0 & 0 & 0 & 0 & 0 & 0 & 0 & 1 & 0 & 0 & 0 & -1 & 0 & 0 & 0 & 0 & 0 & 0 & 0 \\
 0 & 0 & 0 & 0 & 0 & 0 & 0 & 0 & 0 & 0 & 0 & 0 & 0 & 0 & 0 & 0 & 0 & 0 & 0 & 1 & 0 & 0 & -1 & 0 & 0 & 0 & 0 & 0 & 0 & 0 \\
\ea
\right)
$}
~.~
\eea
The toric diagram for the corresponding toric Calabi-Yau 3-fold in \fref{fig_02} is then given by,
\beal{es01a05}
G_t^{A}=
\resizebox{0.37\textwidth}{!}{$
\left(
\ba{cccccc|cc|cccccccccc|cccccccccccc}
  p_1 & p_2 & p_3 & p_4 & p_5 & p_6 & q_1 & q_2 & s_1 & s_2 & s_3 & s_4 & s_5 & s_6 & s_7 & s_8 & s_9 & s_{10} & u_1 & u_2 & u_3 & u_4 & u_5 & u_6 & u_7 & u_8 & u_9 & u_{10} & u_{11} & u_{12}  \\
\hline
-1 & -1 & 0 & 0 & 1 & 1 & 1 & 1 & 
0 & 0 & 0 & 0 & 0 & 0 & 0 & 0 & 0 & 0 & 0 & 0 & 0 & 0 & 0 & 0 & 0 & 0 & 0 & 0 & 0 & 0 \\
 -1 & 0 & -2 & 1 & 1 & -1 & 0 & 0 & 
 -1 & -1 & -1 & -1 & -1 & -1 & -1 & -1 & -1 & -1 & 0 & 0 & 0 & 0 & 0 & 0 & 0 & 0 & 0 & 0 & 0 & 0 \\
 \hline
 1 & 1 & 1 & 1 & 1 & 1 & 1 & 1 & 1 & 1 & 1 & 1 & 1 & 1 & 1 & 1 & 1 & 1 & 1 & 1 & 1 & 1 & 1 & 1 & 1 & 1 & 1 & 1 & 1 & 1 \\
\ea
\right)
$}
~,~
\eea
where the columns correspond to GLSM fields
associated to the vertices of the toric diagram.

The Hilbert series of the mesonic moduli space $\mathcal{M}^{mes}_{A}$ for the abelian $4d$ $\mathcal{N}=1$ theory
takes the following form,
\beal{es01a06}
&&\nn
g(t_a,y_q,y_s,y_u;\mathcal{M}^{mes}_{A})=\frac{P(t_a,y_q,y_s,y_u;\mathcal{M}^{mes}_{A})}{(1-y_s y_u t_1^2 t_2^2 t_3 t_4 )}
\\&& \nn \hspace{0.5cm} \times
\frac{1}{(1-y_q^2 y_s y_u t_3 t_4 t_5^2 t_6^2 )(1-y_q y_s^2 y_u t_1^2 t_2 t_3^3 t_6^2 )}
\\&& \nn \hspace{0.5cm} \times
\frac{1}{(1 - y_q^2 y_s^2 y_u t_1 t_3^3 t_5 t_6^3)(1 - y_q y_s y_u^2 t_1^2 t_2^3 t_4^3 t_5^2)}
\\&&  \hspace{0.5cm} \times
\frac{1}{(1 - y_q^3 y_s y_u^2 t_2 t_4^3 t_5^4 t_6^2)}
~,~
\eea
where $t_a$ is the fugacity corresponding to the GLSM fields $p_a$ 
associated to the extremal vertices of the toric diagram in \fref{fig_02}. 
We also have the fugacities $y_q$, $y_s$ and $y_u$, which correspond to products of GLSM fields
$q_1 q_2$, $s_1 \dots s_{10}$, and $u_1 \dots u_{12}$, respectively, that are associated to non-extremal vertices of the toric diagram. 
The numerator $P(t_a,y_q,y_s,y_u;\mathcal{M}^{mes}_{A})$ of the Hilbert series in \eref{es01a06} is as follows,
\beal{es01a07}
&&\nn
P(t_a,y_q,y_s,y_u;\mathcal{M}^{mes}_{A})=1+y_q y_s y_u t_1 t_2 t_3 t_4 t_5 t_6
\\&& \nn \hspace{0.5cm}
+y_q^2 y_s y_u^2 t_1 t_2^2 t_4^3 t_5^3 t_6-y_q^2 y_s^2 y_u^3 t_1^3 t_2^4 t_3 t_4^4 t_5^3 t_6
\\&& \nn \hspace{0.5cm}
-y_q^2 y_s^3 y_u^3 t_1^4 t_2^4 t_3^3 t_4^3 t_5^2 t_6^2-y_q^3 y_s^2 y_u^3 t_1^2 t_2^3 t_3 t_4^4 t_5^4 t_6^2
\\&& \nn \hspace{0.5cm}
-y_q^2 y_s^3 y_u^2 t_1^3 t_2^2 t_3^4 t_4 t_5 t_6^3-2 y_q^3 y_s^3 y_u^3 t_1^3 t_2^3 t_3^3 t_4^3 t_5^3 t_6^3
\\&& \nn \hspace{0.5cm}   
+y_q^3 y_s^4 y_u^4 t_1^5 t_2^5 t_3^4 t_4^4 t_5^3 t_6^3-y_q^4 y_s^2 y_u^3 t_1 t_2^2 t_3 t_4^4 t_5^5 t_6^3
\\&& \nn \hspace{0.5cm}
-y_q^3 y_s^3 y_u^2 t_1^2 t_2 t_3^4 t_4 t_5^2 t_6^4-2 y_q^4 y_s^3 y_u^3 t_1^2 t_2^2 t_3^3 t_4^3 t_5^4 t_6^4
\\&& \nn \hspace{0.5cm}   
+2 y_q^4 y_s^4 y_u^4 t_1^4 t_2^4 t_3^4 t_4^4 t_5^4 t_6^4+y_q^5 y_s^4 y_u^5 t_1^4 t_2^5 t_3^3 t_4^6 t_5^6 t_6^4
\\&& \nn \hspace{0.5cm}   
+y_q^4 y_s^5 y_u^4 t_1^5 t_2^4 t_3^6 t_4^3 t_5^3 t_6^5-y_q^5 y_s^3 y_u^3 t_1 t_2 t_3^3 t_4^3 t_5^5 t_6^5
\\&& \nn \hspace{0.5cm}   
+2 y_q^5 y_s^4 y_u^4 t_1^3 t_2^3 t_3^4 t_4^4 t_5^5 t_6^5+y_q^6 y_s^4 y_u^5 t_1^3 t_2^4 t_3^3 t_4^6 t_5^7 t_6^5
\\&& \nn \hspace{0.5cm}   
+y_q^5 y_s^5 y_u^4 t_1^4 t_2^3 t_3^6 t_4^3 t_5^4 t_6^6+y_q^6 y_s^4 y_u^4 t_1^2 t_2^2 t_3^4 t_4^4 t_5^6 t_6^6
\\&& \nn \hspace{0.5cm}   
+y_q^6 y_s^5 y_u^4 t_1^3 t_2^2 t_3^6 t_4^3 t_5^5 t_6^7-y_q^6 y_s^6 y_u^5 t_1^5 t_2^4 t_3^7 t_4^4 t_5^5 t_6^7
\\&& \hspace{0.5cm}   
-y_q^7 y_s^6 y_u^6 t_1^5 t_2^5 t_3^6 t_4^6 t_5^7 t_6^7-y_q^8 y_s^7 y_u^7 t_1^6 t_2^6 t_3^7 t_4^7 t_5^8 t_6^8 
~.~
\eea
By setting $t_a=\overline{t}$ for the extremal GLSM fields
and $y_q=y_s=y_u=1$ for all other fugacities, we obtain an unrefined Hilbert series for $\mathcal{M}^{mes}_{A}$, which takes the form,
\beal{es01a08}
&&
P(\overline{t};\mathcal{M}^{mes}_{A})
=
\frac{
(1-\overline{t}^2)^2
}{
(1-\overline{t}^6)(1-\overline{t}^8)^2 (1-\overline{t}^{10})^2
}
\nn\\
&&
\hspace{0.25cm}
\times (1 + 2 \overline{t}^2 + 3 \overline{t}^4 + 6 \overline{t}^6 + 9 \overline{t}^8 + 13 \overline{t}^{10} + 19 \overline{t}^{12} 
\nn\\
&&
\hspace{0.5cm}
+ 23 \overline{t}^{14} + 25 \overline{t}^{16} + 23 \overline{t}^{18} + 19 \overline{t}^{20} + 13 \overline{t}^{22} + 9 \overline{t}^{24} 
\nn\\
&&
\hspace{0.5cm}
+ 6 \overline{t}^{26} + 3 \overline{t}^{28} 
+ 2 \overline{t}^{30} + \overline{t}^{32})
~.~
\eea
The palindromic numerator of the Hilbert series above indicates that $\mathcal{M}^{mes}_{A}$ is Calabi-Yau \cite{stanley1978hilbert}.

The global symmetry of the mesonic moduli space $\mathcal{M}^{mes}_{A}$
for the $4d$ $\mathcal{N}=1$ theory
takes the form
$U(1)_{f_1}\times U(1)_{f_2} \times U(1)_R$, 
where $U(1)_{f_1}\times U(1)_{f_2}$ is the mesonic flavor symmetry and
$U(1)_R$ corresponds to the R-symmetry. 
Using $a$-maximization \cite{Intriligator:2003jj, Butti:2005vn, Butti:2005ps}, we determine the superconformal $U(1)_R$
charges of the chiral fields $X_{ij}$,
which in turn fix the $U(1)_R$ charges $R_a$ of the GLSM fields $p_a$.
Under the AdS/CFT correspondence, these $U(1)_R$ charges also correspond to the minimum volumes 
for divisors $D_a$ of the corresponding toric Calabi-Yau 3-fold. 
\tref{tab_01} and \tref{tab_02} summarize the 
$U(1)_R$ charges on the GLSM fields and the chiral fields of Model A, respectively, 
with the corresponding mesonic flavor symmetry charges.
We note here that the global symmetry charges, including the $U(1)_R$ charges, on chiral fields are specific to Model A, 
whereas the charges on the extremal GLSM fields $p_a$ are the same for Models A and B due to the fact that they correspond to the same toric Calabi-Yau 3-fold.

\subsection{Model B}

\begin{figure}[H]
\begin{center}
\resizebox{0.85\hsize}{!}{
\includegraphics[height=5cm]{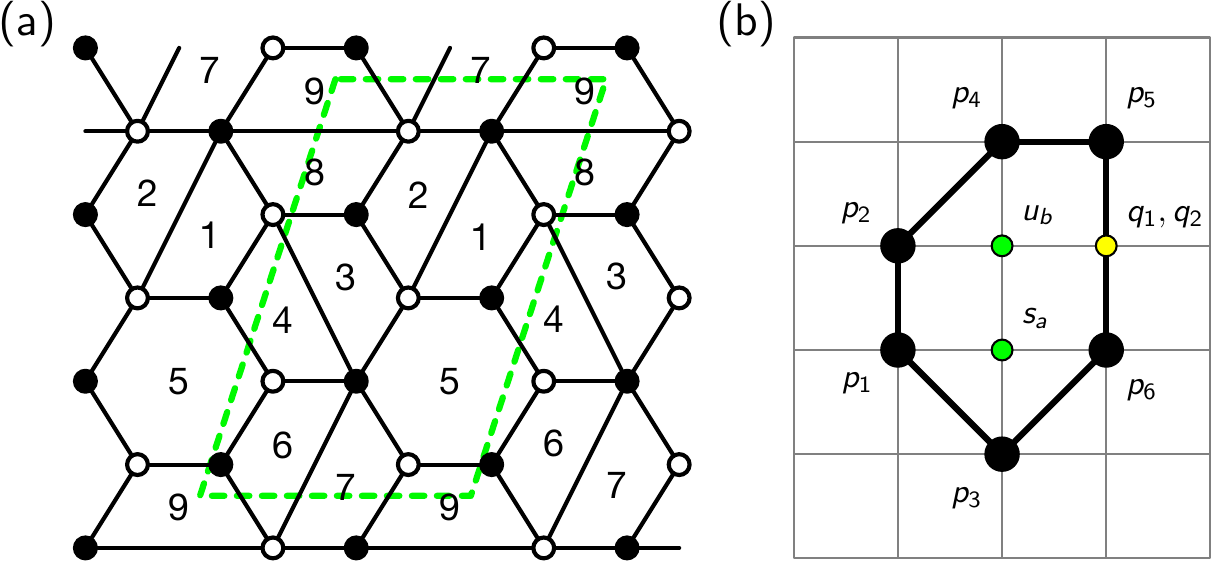} 
}
\caption{
(a) The brane tiling and (b) the labelled toric diagram
for Model B, where
vertices in the toric diagram correspond to GLSM fields $p_a$.
The dotted quadrilateral in the brane tiling indicates the fundamental domain on the 2-torus $T^2$.
\label{f_fig03}}
 \end{center}
 \end{figure}

The quiver for Model B 
is identical to that of Model A, shown in \fref{fig_01}, but the superpotentials differ. The superpotential of Model B is,
\beal{es02a01}
&&
W^{B}=
X_{97} X_{75} X_{59}+X_{54} X_{46} X_{65}+X_{53} X_{32} X_{21} X_{15}
\nn\\
&&
\hspace{0.25cm}
+X_{41}X_{18}X_{83}X_{34}+X_{96}X_{67}X_{72}X_{28} X_{89}
\nn\\
&&
\hspace{0.25cm}
-X_{96} X_{65} X_{59}- X_{54} X_{41} X_{15} - X_{32} X_{28} X_{83} 
\nn\\
&&
\hspace{0.25cm}
- X_{97} X_{72} X_{21} X_{18} X_{89}-X_{75}X_{53}X_{34}X_{46} X_{67}
~,~
\eea
and the corresponding brane tiling is shown in \fref{f_fig03}. 
Applying the forward algorithm, we obtain the perfect matching matrix $P^{B}$ for Model B, given by
\beal{es02a02}
P^{B}=
\resizebox{0.37\textwidth}{!}{$
\left(
\ba{c|cccccc|cc|cccccccccc|cccccccccccc}
\; &  p_1 & p_2 & p_3 & p_4 & p_5 & p_6 & q_1 & q_2 & s_1 & s_2 & s_3 & s_4 & s_5 & s_6 & s_7 & s_8 & s_9 & s_{10} & u_1 & u_2 & u_3 & u_4 & u_5 & u_6 & u_7 & u_8 & u_9 & u_{10} & u_{11} & u_{12}  \\
\hline
 X_{15} &    1 & 0 & 1 & 0 & 0 & 0 & 0 & 0 & 1 & 1 & 0 & 0 & 0 & 0 & 0 & 1 & 1 & 1 & 0 & 0 & 0 & 0 & 1 & 1 & 0 & 0 & 0 & 0 & 0 & 0 \\
 X_{18} &    0 & 0 & 0 & 0 & 1 & 0 & 1 & 0 & 0 & 1 & 0 & 0 & 0 & 0 & 0 & 0 & 0 & 0 & 0 & 0 & 0 & 1 & 0 & 1 & 0 & 0 & 0 & 0 & 0 & 0 \\
 X_{21} &    0 & 0 & 0 & 0 & 0 & 1 & 0 & 1 & 0 & 0 & 1 & 1 & 0 & 0 & 0 & 0 & 0 & 0 & 0 & 0 & 0 & 0 & 0 & 0 & 1 & 0 & 0 & 0 & 0 & 0 \\
 X_{28} &    1 & 1 & 0 & 0 & 0 & 0 & 0 & 0 & 0 & 1 & 1 & 1 & 0 & 0 & 0 & 0 & 0 & 0 & 0 & 0 & 0 & 1 & 0 & 1 & 1 & 0 & 0 & 0 & 0 & 0 \\
 X_{32} &    0 & 0 & 0 & 1 & 1 & 0 & 1 & 0 & 0 & 0 & 0 & 0 & 1 & 1 & 0 & 0 & 0 & 0 & 1 & 1 & 0 & 0 & 0 & 0 & 0 & 1 & 1 & 1 & 1 & 0 \\ 
 X_{34} &    1 & 0 & 0 & 0 & 0 & 0 & 0 & 0 & 0 & 0 & 0 & 1 & 0 & 1 & 0 & 0 & 0 & 0 & 0 & 1 & 0 & 0 & 0 & 0 & 0 & 0 & 0 & 0 & 1 & 0 \\ 
 X_{41} &    0 & 1 & 0 & 1 & 0 & 0 & 0 & 0 & 0 & 0 & 1 & 0 & 1 & 0 & 0 & 0 & 0 & 0 & 1 & 0 & 0 & 0 & 0 & 0 & 1 & 1 & 1 & 1 & 0 & 0 \\ 
 X_{53} &    0 & 1 & 0 & 0 & 0 & 0 & 0 & 0 & 0 & 0 & 0 & 0 & 0 & 0 & 1 & 0 & 0 & 0 & 0 & 0 & 1 & 1 & 0 & 0 & 0 & 0 & 0 & 0 & 0 & 1 \\
 X_{54} &    0 & 0 & 0 & 0 & 1 & 1 & 1 & 1 & 0 & 0 & 0 & 1 & 0 & 1 & 1 & 0 & 0 & 0 & 0 & 1 & 1 & 1 & 0 & 0 & 0 & 0 & 0 & 0 & 1 & 1 \\
 X_{83} &    0 & 0 & 1 & 0 & 0 & 1 & 0 & 1 & 1 & 0 & 0 & 0 & 0 & 0 & 1 & 1 & 1 & 1 & 0 & 0 & 1 & 0 & 1 & 0 & 0 & 0 & 0 & 0 & 0 & 1 \\
 X_{46} &    0 & 0 & 1 & 0 & 0 & 0 & 0 & 0 & 1 & 1 & 1 & 0 & 1 & 0 & 0 & 0 & 0 & 1 & 1 & 0 & 0 & 0 & 0 & 0 & 0 & 0 & 0 & 1 & 0 & 0 \\
 X_{59} &    0 & 0 & 1 & 0 & 0 & 1 & 1 & 0 & 1 & 1 & 1 & 1 & 1 & 1 & 1 & 0 & 0 & 0 & 1 & 1 & 1 & 1 & 0 & 0 & 0 & 0 & 0 & 0 & 0 & 0 \\
 X_{65} &    1 & 1 & 0 & 1 & 0 & 0 & 0 & 0 & 0 & 0 & 0 & 0 & 0 & 0 & 0 & 1 & 1 & 0 & 0 & 0 & 0 & 0 & 1 & 1 & 1 & 1 & 1 & 0 & 0 & 0 \\
 X_{67} &    0 & 0 & 0 & 0 & 0 & 1 & 1 & 0 & 0 & 0 & 0 & 0 & 0 & 0 & 0 & 0 & 1 & 0 & 0 & 0 & 0 & 0 & 0 & 0 & 0 & 0 & 1 & 0 & 0 & 0 \\ 
 X_{72} &    0 & 0 & 1 & 0 & 0 & 0 & 0 & 0 & 0 & 0 & 0 & 0 & 1 & 1 & 1 & 1 & 0 & 0 & 0 & 0 & 0 & 0 & 0 & 0 & 0 & 1 & 0 & 0 & 0 & 0 \\ 
 X_{75} &    0 & 0 & 0 & 1 & 1 & 0 & 0 & 1 & 0 & 0 & 0 & 0 & 0 & 0 & 0 & 1 & 0 & 0 & 0 & 0 & 0 & 0 & 1 & 1 & 1 & 1 & 0 & 0 & 0 & 0 \\
 X_{89} &    0 & 0 & 0 & 1 & 0 & 0 & 0 & 0 & 1 & 0 & 0 & 0 & 0 & 0 & 0 & 0 & 0 & 0 & 1 & 1 & 1 & 0 & 1 & 0 & 0 & 0 & 0 & 0 & 0 & 0 \\
 X_{97} &    1 & 1 & 0 & 0 & 0 & 0 & 0 & 0 & 0 & 0 & 0 & 0 & 0 & 0 & 0 & 0 & 1 & 1 & 0 & 0 & 0 & 0 & 0 & 0 & 0 & 0 & 1 & 1 & 1 & 1 \\
 X_{96} &    0 & 0 & 0 & 0 & 1 & 0 & 0 & 1 & 0 & 0 & 0 & 0 & 0 & 0 & 0 & 0 & 0 & 1 & 0 & 0 & 0 & 0 & 0 & 0 & 0 & 0 & 0 & 1 & 1 & 1 \\
\ea
\right)
$}
~,~
\eea
which is manifestly different from the perfect matching matrix for Model A in \eref{es01a02}, even though the two models share the same set of chiral fields $X_{ij}$.
The associated F-term charge matrix $Q_F^{B}$ is given by, 
\beal{es02a03}
Q_F^{B}=
\resizebox{0.37\textwidth}{!}{$
\left(
\ba{cccccc|cc|cccccccccc|cccccccccccc}
  p_1 & p_2 & p_3 & p_4 & p_5 & p_6 & q_1 & q_2 & s_1 & s_2 & s_3 & s_4 & s_5 & s_6 & s_7 & s_8 & s_9 & s_{10} & u_1 & u_2 & u_3 & u_4 & u_5 & u_6 & u_7 & u_8 & u_9 & u_{10} & u_{11} & u_{12}  \\
\hline
 1 & 0 & 0 & 0 & 0 & 0 & 0 & 1 & 0 & 0 & 0 & 0 & 0 & 0 & 0 & 0 & 0 & -1 & 0 & 0 & 0 & 0 & 0 & 0 & -1 & 0 & 0 & 1 & -1 & 0 \\
 0 & 1 & 0 & 0 & 0 & 0 & 0 & 1 & 0 & 0 & 0 & 0 & 0 & 0 & 0 & 0 & 0 & 0 & 0 & 0 & 0 & 0 & 0 & 0 & -1 & 0 & 0 & 0 & 0 & -1 \\
 0 & 0 & 1 & 0 & 0 & 0 & 0 & 0 & 0 & 0 & 0 & 0 & 0 & 0 & 0 & 0 & 0 & -2 & 0 & 0 & 0 & -1 & 0 & 1 & 0 & -1 & 0 & 1 & 0 & 1 \\
 0 & 0 & 0 & 1 & 0 & 0 & 0 & 0 & 0 & 0 & 0 & 0 & 0 & 0 & 0 & 0 & 0 & 1 & 0 & 0 & 0 & 0 & -1 & 0 & 0 & 0 & 0 & -1 & 0 & 0 \\
 0 & 0 & 0 & 0 & 1 & 0 & 0 & -1 & 0 & 0 & 0 & 0 & 0 & 0 & 0 & 0 & 0 & 1 & 0 & 0 & 0 & 0 & 0 & -1 & 1 & 0 & 0 & -1 & 0 & 0 \\
 0 & 0 & 0 & 0 & 0 & 1 & 0 & -1 & 0 & 0 & 0 & 0 & 0 & 0 & 0 & 0 & 0 & -1 & 0 & 0 & 0 & -1 & 0 & 1 & 0 & 0 & -1 & 1 & 0 & 1 \\
 0 & 0 & 0 & 0 & 0 & 0 & 1 & -1 & 0 & 0 & 0 & 0 & 0 & 0 & 0 & 0 & 0 & 0 & 0 & 0 & 0 & -1 & 0 & 0 & 1 & 0 & -1 & 0 & 0 & 1 \\
 0 & 0 & 0 & 0 & 0 & 0 & 0 & 0 & 1 & 0 & 0 & 0 & 0 & 0 & 0 & 0 & 0 & -1 & 0 & 0 & 0 & -1 & -1 & 1 & 0 & 0 & 0 & 0 & 0 & 1 \\
 0 & 0 & 0 & 0 & 0 & 0 & 0 & 0 & 0 & 1 & 0 & 0 & 0 & 0 & 0 & 0 & 0 & -1 & 0 & 0 & 0 & -1 & 0 & 0 & 0 & 0 & 0 & 0 & 0 & 1 \\
 0 & 0 & 0 & 0 & 0 & 0 & 0 & 0 & 0 & 0 & 1 & 0 & 0 & 0 & 0 & 0 & 0 & -1 & 0 & 0 & 0 & -1 & 0 & 1 & -1 & 0 & 0 & 0 & 0 & 1 \\
 0 & 0 & 0 & 0 & 0 & 0 & 0 & 0 & 0 & 0 & 0 & 1 & 0 & 0 & 0 & 0 & 0 & -1 & 0 & 0 & 0 & -1 & 0 & 1 & -1 & 0 & 0 & 1 & -1 & 1 \\
 0 & 0 & 0 & 0 & 0 & 0 & 0 & 0 & 0 & 0 & 0 & 0 & 1 & 0 & 0 & 0 & 0 & -1 & 0 & 0 & 0 & -1 & 0 & 1 & 0 & -1 & 0 & 0 & 0 & 1 \\
 0 & 0 & 0 & 0 & 0 & 0 & 0 & 0 & 0 & 0 & 0 & 0 & 0 & 1 & 0 & 0 & 0 & -1 & 0 & 0 & 0 & -1 & 0 & 1 & 0 & -1 & 0 & 1 & -1 & 1 \\
 0 & 0 & 0 & 0 & 0 & 0 & 0 & 0 & 0 & 0 & 0 & 0 & 0 & 0 & 1 & 0 & 0 & -1 & 0 & 0 & 0 & -1 & 0 & 1 & 0 & -1 & 0 & 1 & 0 & 0 \\
 0 & 0 & 0 & 0 & 0 & 0 & 0 & 0 & 0 & 0 & 0 & 0 & 0 & 0 & 0 & 1 & 0 & -1 & 0 & 0 & 0 & 0 & 0 & 0 & 0 & -1 & 0 & 1 & 0 & 0 \\
 0 & 0 & 0 & 0 & 0 & 0 & 0 & 0 & 0 & 0 & 0 & 0 & 0 & 0 & 0 & 0 & 1 & -1 & 0 & 0 & 0 & 0 & 0 & 0 & 0 & 0 & -1 & 1 & 0 & 0 \\
 0 & 0 & 0 & 0 & 0 & 0 & 0 & 0 & 0 & 0 & 0 & 0 & 0 & 0 & 0 & 0 & 0 & 0 & 1 & 0 & 0 & -1 & -1 & 1 & 0 & 0 & 0 & -1 & 0 & 1 \\
 0 & 0 & 0 & 0 & 0 & 0 & 0 & 0 & 0 & 0 & 0 & 0 & 0 & 0 & 0 & 0 & 0 & 0 & 0 & 1 & 0 & -1 & -1 & 1 & 0 & 0 & 0 & 0 & -1 & 1 \\
 0 & 0 & 0 & 0 & 0 & 0 & 0 & 0 & 0 & 0 & 0 & 0 & 0 & 0 & 0 & 0 & 0 & 0 & 0 & 0 & 1 & -1 & -1 & 1 & 0 & 0 & 0 & 0 & 0 & 0 \\
\ea
\right)
$}
~,~
\eea
and
the D-term charge matrix $Q_D^{B}$ is given by,
\beal{es02a04}
Q_D^{B}=
\resizebox{0.37\textwidth}{!}{$
\left(
\ba{cccccc|cc|cccccccccc|cccccccccccc}
  p_1 & p_2 & p_3 & p_4 & p_5 & p_6 & q_1 & q_2 & s_1 & s_2 & s_3 & s_4 & s_5 & s_6 & s_7 & s_8 & s_9 & s_{10} & u_1 & u_2 & u_3 & u_4 & u_5 & u_6 & u_7 & u_8 & u_9 & u_{10} & u_{11} & u_{12}  \\
\hline
0 & 0 & 1 & 0 & 0 & 0 & 0 & 0 & 0 & 0 & 0 & 0 & 0 & 0 & 0 & 0 & -2 & 0 & 0 & 0 & 0 & 0 & 1 & 0 & 0 & 0 & 0 & 0 & 0 & 0 \\
0 & 0 & 0 & 0 & 1 & 0 & -1 & 0 & 0 & 0 & 0 & 0 & 0 & 0 & 0 & 0 & 1 & 0 & 0 & 0 & 0 & 0 & -1 & 0 & 0 & 0 & 0 & 0 & 0 & 0 \\
0 & 0 & 0 & 0 & 0 & 0 & 0 & 0 & 1 & 0 & 0 & 0 & 0 & 0 & 0 & 0 & -1 & 0 & 0 & 0 & 0 & 0 & 0 & 0 & 0 & 0 & 0 & 0 & 0 & 0 \\
0 & 0 & 0 & 0 & 0 & 0 & 0 & 0 & 0 & 1 & 0 & 0 & 0 & 0 & 0 & 0 & -1 & 0 & 0 & 0 & 0 & 0 & 0 & 0 & 0 & 0 & 0 & 0 & 0 & 0 \\
0 & 0 & 0 & 0 & 0 & 0 & 0 & 0 & 0 & 0 & 1 & 0 & 0 & 0 & 0 & 0 & -1 & 0 & 0 & 0 & 0 & 0 & 0 & 0 & 0 & 0 & 0 & 0 & 0 & 0 \\
0 & 0 & 0 & 0 & 0 & 0 & 0 & 0 & 0 & 0 & 0 & 0 & 0 & 0 & 0 & 0 & 0 & 0 & 1 & 0 & 0 & 0 & -1 & 0 & 0 & 0 & 0 & 0 & 0 & 0 \\
0 & 0 & 0 & 0 & 0 & 0 & 0 & 0 & 0 & 0 & 0 & 0 & 0 & 0 & 0 & 0 & 0 & 0 & 0 & 1 & 0 & 0 & -1 & 0 & 0 & 0 & 0 & 0 & 0 & 0 \\
0 & 0 & 0 & 0 & 0 & 0 & 0 & 0 & 0 & 0 & 0 & 0 & 0 & 0 & 0 & 0 & 0 & 0 & 0 & 0 & 1 & 0 & -1 & 0 & 0 & 0 & 0 & 0 & 0 & 0 \\
\ea
\right)
$}
~.~
\eea
By taking the kernel of the combined charge matrix $Q_t^{B}$, 
we obtain,
\beal{es02a05}
G_t^{B}=
\resizebox{0.37\textwidth}{!}{$
\left(
\ba{cccccc|cc|cccccccccc|cccccccccccc}
  p_1 & p_2 & p_3 & p_4 & p_5 & p_6 & q_1 & q_2 & s_1 & s_2 & s_3 & s_4 & s_5 & s_6 & s_7 & s_8 & s_9 & s_{10} & u_1 & u_2 & u_3 & u_4 & u_5 & u_6 & u_7 & u_8 & u_9 & u_{10} & u_{11} & u_{12}  \\
\hline
-1 & -1 & 0 & 0 & 1 & 1 & 1 & 1 & 
0 & 0 & 0 & 0 & 0 & 0 & 0 & 0 & 0 & 0 & 0 & 0 & 0 & 0 & 0 & 0 & 0 & 0 & 0 & 0 & 0 & 0 \\
 -1 & 0 & -2 & 1 & 1 & -1 & 0 & 0 & 
 -1 & -1 & -1 & -1 & -1 & -1 & -1 & -1 & -1 & -1 & 0 & 0 & 0 & 0 & 0 & 0 & 0 & 0 & 0 & 0 & 0 & 0 \\
 \hline
 1 & 1 & 1 & 1 & 1 & 1 & 1 & 1 & 1 & 1 & 1 & 1 & 1 & 1 & 1 & 1 & 1 & 1 & 1 & 1 & 1 & 1 & 1 & 1 & 1 & 1 & 1 & 1 & 1 & 1 \\
\ea
\right)
$}
~,~
\eea
which contains as columns the coordinates of the vertices 
in the toric diagram for Model B in \fref{f_fig03}.

\begin{table}[ht!]
\centering
\begin{tabular}{|c|c|c|l|}
\hline
\; & $U(1)_{f_1}$ & $U(1)_{f_2}$ & $U(1)_R$ \\
\hline \hline
$X_{15}$ &  0        &  +1	& $R_1+R_3\simeq 0.659722$  \\
$X_{18}$ &  0        &  -1	& $R_5\simeq 0.404904$  \\
$X_{21}$ &  0        &  -1	& $R_6\simeq 0.361689$  \\
$X_{28}$ &  +2        &  0	& $R_1+R_2\simeq 0.594011$  \\
$X_{32}$ &  -1        &  0	& $R_4+R_5\simeq 0.658128$  \\
$X_{34}$ &  +1        &  0	& $R_1\simeq 0.27355$  \\
$X_{41}$ &  0        &  +1	& $R_2+R_4\simeq 0.573685$  \\
$X_{53}$ &  +1        &  0	& $R_2\simeq 0.320461$  \\
$X_{54}$ &  0        &  -2	& $R_5+R_6\simeq 0.766593$  \\
$X_{83}$ &  -1        &  0	& $R_3+R_6\simeq 0.74786$  \\
\hline
$X_{46}$ &  -1        &  +1	& $R_3\simeq 0.386171$  \\
$X_{59}$ &  -1        &  0	& $R_3+R_6\simeq 0.747860$  \\
$X_{65}$ &  +1        &  +1	& $R_1+R_2+R_4\simeq 0.847235$  \\
$X_{67}$ &  0        &  -1	& $R_6\simeq 0.361689$  \\
$X_{72}$ & -1        &  +1 	& $R_3\simeq 0.386171$  \\
$X_{75}$ & -1        &  0 	& $R_4+R_5\simeq 0.658128$\\
$X_{89}$ &  -1        &  +1	& $R_4\simeq 0.253224$  \\
$X_{96}$ &  0 	     &  -1	& $R_5\simeq 0.404904$  \\
$X_{97}$ &  +2	     &  0	& $R_1+R_2\simeq 0.594011 $  \\
\hline
\end{tabular}
\caption{
The charges under the global symmetry on chiral fields in Model B, including the $U(1)_R$ charges that are expressed in 
terms of $U(1)_R$ charges $R_a$ corresponding to the extremal GLSM fields $p_a$.
Note that the global symmetry charges of the chiral fields $X_{46}, \dots, X_{97}$ are identical to those in Model A, whereas the charges of the remaining chiral fields -- those affected by the `tilting’ mutation of the brane tiling -- differ between the two models (up to a re-ordering of charges).
 \label{tab_03}}
\end{table}

We note that the toric diagrams extracted from the matrices $G_t^A$ and $G_t^B$ are identical up to a $GL(2,\mathbb{Z})$ transformation. 
This indicates that the mesonic moduli spaces $\mathcal{M}^{mes}$ associated with the two brane tilings are the same. 
The same conclusion holds for the non-abelian $4d$ $\mathcal{N}=1$ theories with $U(N)$ gauge groups, 
for which the mesonic moduli space is simply the $N$-th symmetric product of $\mathcal{M}^{mes}$.
Moreover, upon computing the Hilbert series of $\mathcal{M}^{mes}$ for Model B, we find that it coincides with that of Model A in \eref{es01a06}, 
\beal{es02a10}
g(t_a,y_q,y_s,y_u;\mathcal{M}^{mes}_{A})
=
g(t_a,y_q,y_s,y_u;\mathcal{M}^{mes}_{B})
~.~
\eea
Since Models A and B share the same set of GLSM fields,
and hence the same set of toric diagram vertices with the same coordinates, 
we use the same set of fugacities $(t_a,y_q,y_s,y_u)$ for both Hilbert series.

Because the mesonic moduli spaces are identical, the global symmetry is also the same for both brane tilings, namely
$U(1)_{f_1} \times U(1)_{f_2} \times U(1)_R$.
As a result, the shared GLSM fields $p_a$ carry the same charges under the global symmetry in the two models. 
This is manifest, for example, in the superconformal $U(1)_R$ charges $R_a$ of the GLSM fields obtained via volume minimization \cite{Martelli:2006yb, Martelli:2005tp}, 
where the $U(1)_R$ charges $R_a$ correspond to the minimal volumes of divisors $D_a$  \cite{Butti:2006au, Bao:2024nyu} in the associated toric Calabi-Yau 3-fold. 
\tref{tab_01} summarizes for both Models A and B the global symmetry charges on GLSM fields $p_a$, including the $U(1)_R$ charges $R_a$.

\begin{table*}[htt!]
\centering
\resizebox{1\textwidth}{!}{
\begin{tabular}{|c|c|c|}
\hline
Generator (GLSM) & Generator (Model A chiral fields) & Generator (Model B chiral fields) 
\\
\hline\hline
$p_1^2 p_2^2 p_3 p_4 q s u$ 
&  
\begin{tabular}{c}$
X_{54} X_{46} X_{65} 
= X_{21} X_{18} X_{83} X_{32}
= X_{97} X_{72}X_{21} X_{18} X_{89}
$\end{tabular}  
&
\begin{tabular}{c}$
X_{53} X_{34} X_{41} X_{15} 
= X_{53} X_{34} X_{46} X_{65} 
= X_{97} X_{72} X_{28} X_{89}
$\end{tabular}
\\
\hline
$p_1 p_2 p_3 p_4 p_5 p_6 q s u$
&
\begin{tabular}{c}$
X_{97} X_{75} X_{59}
= X_{96} X_{65} X_{59}
= X_{54} X_{41} X_{15}
$ \\ $  
= X_{28} X_{83} X_{32}
= X_{53} X_{34} X_{46} X_{65} 
= X_{41} X_{18} X_{83} X_{34} 
$ \\ $  
= X_{75} X_{54} X_{46} X_{67}  
= X_{53} X_{32} X_{21} X_{15} 
 $ \\ $  
 = X_{97} X_{72} X_{28} X_{89}
= X_{96} X_{67} X_{72} X_{21} X_{18} X_{89} 
$\end{tabular} 
&
\begin{tabular}{c}$
X_{97} X_{75} X_{59}
= X_{96} X_{65} X_{59} 
= X_{54} X_{41} X_{15}
$ \\ $  
= X_{54} X_{46} X_{65}
= X_{28} X_{83} X_{32}
= X_{53} X_{32} X_{21} X_{15} 
$ \\ $  
= X_{41} X_{18} X_{83} X_{34} 
= X_{75} X_{53} X_{34} X_{46} X_{67} 
$ \\ $  
= X_{96} X_{67} X_{72} X_{28} X_{89} 
= X_{97} X_{72} X_{21} X_{18} X_{89} 
$\end{tabular}
\\
\hline
$p_3 p_4 p_5^2 p_6^2 q^2 s u$
&
\begin{tabular}{c}$
X_{96} X_{67} X_{75} X_{59} 
= X_{53} X_{34} X_{41} X_{15} 
$ \\ $  
= X_{75} X_{53} X_{34} X_{46} X_{67} 
= X_{96} X_{67} X_{72} X_{28} X_{89} 
$\end{tabular}
&
\begin{tabular}{c}$
X_{96} X_{67} X_{75} X_{59} 
= X_{75} X_{54} X_{46} X_{67}
$ \\ $  
= X_{21} X_{18} X_{83} X_{32}
= X_{96} X_{67} X_{72} X_{21} X_{18} X_{89} 
$\end{tabular}
\\
\hline
$p_1^2 p_2 p_3^3 p_6^2 q s^2 u$ 
& 
\begin{tabular}{c}$
X_{97} X_{72} X_{21} X_{15} X_{59}
= X_{72} X_{21} X_{15} X_{54} X_{46} X_{67}
$ \\ $  
= X_{72} X_{21} X_{18} X_{83} X_{34} X_{46}  X_{67} 
$\end{tabular}
&
\begin{tabular}{c}$
X_{97} X_{72} X_{21} X_{15} X_{59}
= X_{72} X_{28} X_{83} X_{34} X_{46} X_{67} 
$ \\ $  
= X_{72} X_{21} X_{15} X_{53}  X_{34} X_{46} X_{67} 
$ \end{tabular} 
\\
\hline
$p_1^2 p_2^3 p_4^3 p_5^2 q s u^2$
&
\begin{tabular}{c}$
X_{97} X_{75} X_{54} X_{41} X_{18} X_{89} 
= X_{96} X_{65} X_{54} X_{41} X_{18} X_{89} 
$ \\ $  
= X_{97} X_{75} X_{53} X_{32} X_{21} X_{18} X_{89} 
= X_{96} X_{65} X_{53} X_{32} X_{21} X_{18} X_{89} 
$ \end{tabular}
&
\begin{tabular}{c}$
X_{97} X_{75} X_{53}  X_{32} X_{28} X_{89} 
= X_{96} X_{65}  X_{53} X_{32} X_{28} X_{89}
$ \\ $  
= X_{97} X_{75} X_{53} X_{34} X_{41} X_{18}  X_{89}
= X_{96} X_{65} X_{53} X_{34} X_{41} X_{18} X_{89} 
$ \end{tabular} 
\\
\hline
$p_1 p_3^3 p_5 p_6^3 q^2 s^2 u$
&
\begin{tabular}{c}$
X_{96} X_{67} X_{72} X_{21} X_{15} X_{59} 
= X_{72} X_{28} X_{83} X_{34} X_{46} X_{67} 
$ \\ $  
= X_{72} X_{21} X_{15} X_{53} X_{34} X_{46} X_{67} 
$\end{tabular}
&
\begin{tabular}{c}$
X_{96}  X_{67} X_{72} X_{21} X_{15} X_{59} 
= X_{72}  X_{21}  X_{15} X_{54} X_{46} X_{67} 
$ \\ $  
= X_{72}  X_{21} X_{18} X_{83} X_{34} X_{46} X_{67}
$\end{tabular}
\\
\hline
$p_1 p_2^2 p_4^3 p_5^3 p_6 q^2 s u^2$
&
\begin{tabular}{c}$
X_{97} X_{75} X_{53} X_{32} X_{28} X_{89} 
= X_{96} X_{65} X_{53} X_{32} X_{28} X_{89}
$ \\ $  
= X_{97} X_{75} X_{53} X_{34} X_{41} X_{18} X_{89} 
= X_{96} X_{67} X_{75} X_{54} X_{41} X_{18} X_{89} 
$ \\ $  
= X_{96} X_{65} X_{53} X_{34} X_{41} X_{18} X_{89} 
= X_{96} X_{67} X_{75} X_{53} X_{32} X_{21} X_{18} X_{89} 
$\end{tabular}
&
\begin{tabular}{c}$
X_{97} X_{75} X_{54} X_{41} X_{18} X_{89}
= X_{96} X_{65}  X_{54} X_{41} X_{18} X_{89}
$ \\ $  
= X_{97} X_{75} X_{53} X_{32} X_{21} X_{18} X_{89} 
= X_{96} X_{67} X_{75} X_{53} X_{32} X_{28} X_{89} 
$ \\ $  
= X_{96} X_{65} X_{53} X_{32} X_{21} X_{18}  X_{89}
= X_{96} X_{67} X_{75} X_{53} X_{34} X_{41} X_{18} X_{89}
$\end{tabular}
\\
\hline
$p_2 p_4^3 p_5^4 p_6^2 q^3 s u^2$
&
\begin{tabular}{l}$
X_{96} X_{67} X_{75} X_{53} X_{32} X_{28} X_{89} 
= X_{96} X_{67} X_{75} X_{53} X_{34} X_{41} X_{18} X_{89} 
$\end{tabular}
&
\begin{tabular}{l}$
X_{96} X_{67} X_{75} X_{54} X_{41} X_{18} X_{89} 
= X_{96} X_{67} X_{75} X_{53} X_{32} X_{21} X_{18} X_{89} 
$\end{tabular} 
\\
\hline
\end{tabular}
}
\caption{
The generators of the shared mesonic moduli space $\mathcal{M}^{mes}$ are given both in terms of the GLSM fields $p_a$ (which are common to Models A and B) and in terms of the chiral fields $X_{ij}$ of the shared quiver. 
Since the decomposition of the GLSM fields into chiral fields differs between the two models, the same set of generators is realized differently in terms of the same chiral field content.
 \label{tab_04}}
\end{table*}

Although the two $4d$ $\mathcal{N}=1$ theories share the same quiver and hence the same set of chiral fields $X_{ij}$,
the global symmetry charges of the chiral fields are not necessarily the same, since the two models have different superpotentials. 
This is most evident when we calculate the superconformal $U(1)_R$ charges carried by 
the chiral fields via $a$-maximization \cite{Intriligator:2003jj, Butti:2005vn, Butti:2005ps} and express them in terms of the $U(1)_R$ charges on GLSM fields $R_a$ shared between the two models, as summarized in \tref{tab_03}.
In particular, the $U(1)_R$ charges of the chiral fields in Model B, which are affected by the
`tilting' mutation on the brane tiling,
differ from those of the same chiral fields in Model A as shown in \tref{tab_02}.

We also obtain the plethystic logarithm \cite{Benvenuti:2006qr, Feng:2007ur} of the shared Hilbert series in $\eref{es01a06}$, which takes the form,
\beal{es03a01}
&&
\text{PL}[g(t_a,y_q,y_s,y_u;\mathcal{M}^{mes})]
=
t_1^2 t_2^2 t_3 t_4 y_s y_u  
\nn\\
&& 
\hspace{0.25cm}
+ t_1 t_2 t_3 t_4 t_5 t_6 y_q y_s y_u +t_3 t_4 t_5^2 t_6^2 y_q^2 y_s y_u  
\nn\\
&& 
\hspace{0.25cm}
+ t_1^2 t_2 t_3^3 t_6^2 y_q y_s^2 y_u + t_1^2 t_2^3 t_4^3 t_5^2 y_q y_s y_u^2 
\nn\\
&& 
\hspace{0.25cm}
+ t_1 t_3^3 t_5 t_6^3 y_q^2 y_s^2 y_u + t_1 t_2^2 t_4^3 t_5^3 t_6 y_q^2 y_s y_u^2 
\nn\\
&& 
\hspace{0.25cm}
+ t_2 t_4^3 t_5^4 t_6^2 y_q^3 y_s y_u^2 -(t_1^2 t_2^4 t_4^6 t_5^6 t_6^2 y_q^4 y_s^2 y_u^4
\nn\\
&& 
\hspace{0.25cm}
+t_1^2 t_2^2 t_3^2 t_4^2 t_5^2 t_6^2 y_q^2 y_s^2 y_u^2  
+t_1^3 t_2^2 t_3^4 t_4 t_5 t_6^3 y_q^2 y_s^3 y_u^2
\nn\\
&& 
\hspace{0.25cm}
+t_1^3 t_2^4 t_3 t_4^4 t_5^3 t_6 y_q^2 y_s^2 y_u^3+
t_1^2 t_2 t_3^4 t_4 t_5^2 t_6^4 y_q^3 y_s^3 y_u^2 
\nn\\
&& 
\hspace{0.25cm}
+ 2 t_1^2 t_2^3 t_3 t_4^4 t_5^4 t_6^2 y_q^3 y_s^2 y_u^3 + 
t_1^4 t_2^4 t_3^3 t_4^3 t_5^2 t_6^2 y_q^2 y_s^3 y_u^3 
\nn\\
&& 
\hspace{0.25cm}
+t_1 t_2^2 t_3 t_4^4 t_5^5 t_6^3 y_q^4 y_s^2 y_u^3 + 
 2 t_1^3 t_2^3 t_3^3 t_4^3 t_5^3 t_6^3 y_q^3 y_s^3 y_u^3 
\nn\\
&& 
\hspace{0.25cm}
+2 t_1^2 t_2^2 t_3^3 t_4^3 t_5^4 t_6^4 y_q^4 y_s^3 y_u^3-t_1 t_2 t_3^3 t_4^3 t_5^5 t_6^5 y_q^5 y_s^3 y_u^3) +
\dots
~,~
\nn\\
\eea
where the non-terminating expansion indicates that the toric Calabi-Yau 3-fold is not a complete intersection. 
The generators of the mesonic moduli space $\mathcal{M}^{mes}$ can be extracted from the leading positive terms in \eref{es03a01}, 
while the subsequent negative terms encode relations among these generators. 
Recalling that the fugacities $t_a$ correspond to extremal GLSM fields $p_a$, 
and that $y_q,y_s,y_u$ correspond to products of non-extremal GLSM fields, 
we identify the generators in terms of GLSM fields as summarized in \tref{tab_04}.

Finally, although the set of GLSM fields is the same for both models, their expressions in terms of chiral fields differ, 
as encoded by the distinct $P$-matrices in \eref{es01a02} and \eref{es02a02}. 
Accordingly, even though Models A and B share the same set of generators, the same GLSM fields $p_a$, and the same chiral fields $X_{ij}$ as dictated by the common quiver in \fref{fig_01}, 
the generators are realized differently in terms of chiral fields in the two models, as shown in \tref{tab_04}.

\section{`Tilting' Mutation and Zig-Zag Paths}

\begin{figure*}[htt!!]
\begin{center}
\resizebox{0.8\hsize}{!}{
\includegraphics[height=5cm]{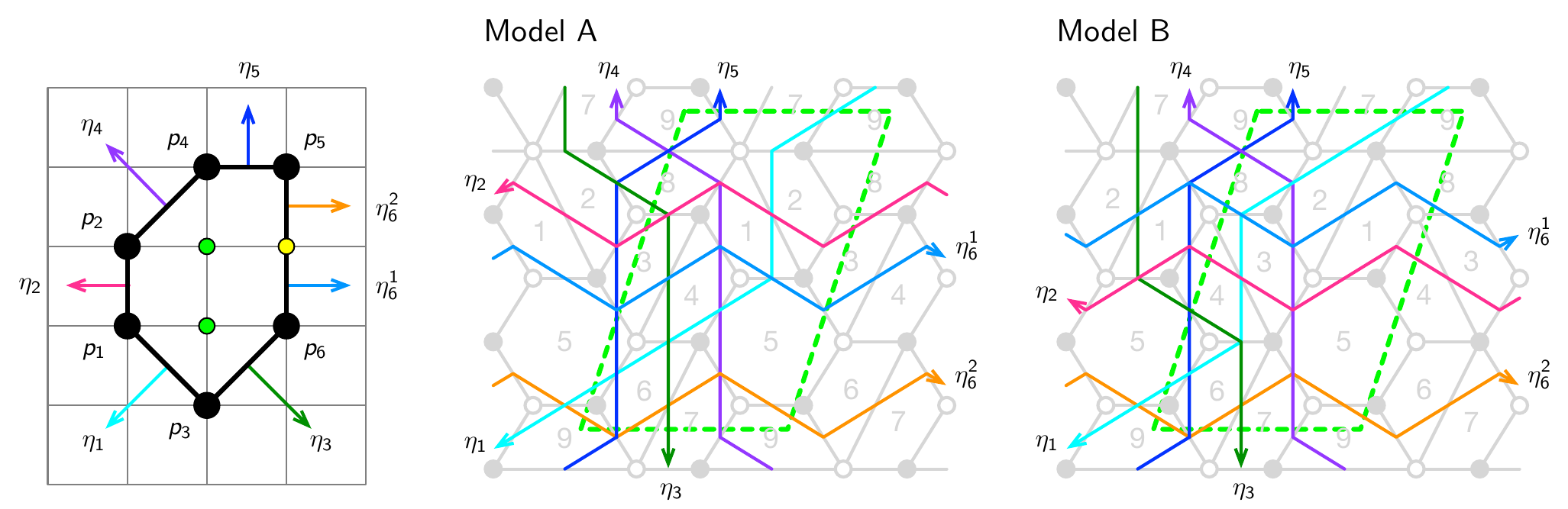} 
}
\caption{
Outward-pointing normal vectors to the boundary edges of the toric diagram
correspond to zig-zag paths $\eta_m$ in the brane tiling. 
The winding number
$w(\eta_m)\in\mathbb{Z}^2$ of a zig-zag path on the 2-torus $T^2$ is given by the direction of the associated normal vector.
The local `tilting' mutation changes the intersection structure of the zig-zag paths while preserving their winding numbers.
In addition, 3 parallel zig-zag paths $\eta_2$, $\eta_6^1$ and $\eta_6^2$ (with $\eta_6^1$ and $\eta_6^2$ having the same direction) exchange their relative ordering under the mutation.
\label{f_fig05}}
 \end{center}
 \end{figure*}

The correspondence between Models A and B is realized by a `tilting' mutation along diagonals of hexagonal faces in the brane tiling, as illustrated in \fref{f_fig04}. 
These diagonals divide each hexagon into two quadrilateral faces in the brane tiling. 
The `tilting' mutation also affects the associated periodic quivers, which are dual to the bipartite graphs of the brane tilings. 
Since periodic quivers on $T^2$ encode not only the quiver data of the associated $4d$ $\mathcal{N}=1$ theories but also their superpotentials, the periodic quivers differ between the two brane tilings. 
Nevertheless, a chosen fundamental domain on $T^2$ contains the invariant quiver shared by both brane tilings.

Brane tilings also admit oriented paths along edges that turn clockwise at white nodes and anti-clockwise at black nodes. 
These are the zig-zag paths $\eta_m$ \cite{Hanany:2005ss, 2009arXiv0901.4662B}. 
Their winding numbers $w(\eta_m)\in\mathbb{Z}^2$ on $T^2$ determine the directions of the outward-pointing normal vectors to the boundary edges of the toric diagram of the corresponding toric Calabi-Yau 3-fold, as illustrated in \fref{f_fig05}. 
Since the mutation preserves the mesonic moduli space of the $4d$ $\mathcal{N}=1$ theories -- 
and hence the underlying toric Calabi-Yau 3-fold 
-- the set of zig-zag paths and their winding numbers in $\mathbb{Z}^2$
remain unchanged. 
We observe, however, that the mutation modifies the intersection pattern among zig-zag paths, 
as shown in \fref{f_fig05}. 
In addition, 3 parallel zig-zag paths $\eta_2$, $\eta_6^1$ and $\eta_6^2$, corresponding to parallel normal vectors in the toric diagram, 
permute under the `tilting' mutation.

Overall, the `tilting' mutation relating Models A and B corresponds to a sequence of Seiberg dualities realized by spider moves of the form shown in \fref{fig_00}. 
It must involve two or more Seiberg dualities, since Models A and B share the same quiver but define distinct $4d$ $\mathcal{N}=1$ theories. 
For the pair discussed here, the single-step `tilting' mutation can be decomposed, for example, into two consecutive Seiberg dualities acting on the gauge groups associated with brane tiling faces 1 and 5.
We note, however, that the `tilting' mutation preserves the quiver together with its node labels, 
whereas the sequence of Seiberg dualities preserves only the shape of the quiver. 
To make the sequence of Seiberg dualities equivalent to the `tilting’ mutation on the brane tiling, 
one must therefore perform an appropriate relabelling of the gauge nodes in the quiver.

\section{Discussion}

In this work, we have shown that two brane tilings describing distinct $4d$ $\mathcal{N}=1$ theories can correspond to the same toric Calabi-Yau 3-fold 
while sharing the same quiver and chiral field content. 
The only distinction between the two theories lies in their superpotentials, 
which are built from the same set of chiral fields. 
Since the two brane tilings have the same mesonic moduli space and share the same quiver, 
the corresponding $4d$ $\mathcal{N}=1$ theories are related by a sequence of Seiberg dualities acting on different gauge groups that, remarkably, 
leaves the quiver invariant. 
We identify this sequence of Seiberg dualities with a single local `tilting' mutation along diagonals of hexagonal faces in the brane tilings. 
In terms of zig-zag paths, the `tilting' mutation preserves not only the mesonic moduli space 
but also the set of zig-zag paths and their winding numbers, which correspond to the $\mathbb{Z}^2$ directions of outward-pointing 
normal vectors along the boundary edges of the shared toric diagram.
Since these outward-pointing normal vectors correspond to legs of a dual $(p,q)$-web diagram
defining a $5d$ $\mathcal{N}=1$ gauge theory \cite{Franco:2023flw, Arias-Tamargo:2024fjt, Franco:2023mkw,CarrenoBolla:2024fxy},
whereas our invariant quiver of the mutated brane tiling can be associated with a $5d$ BPS quiver \cite{Closset:2019juk},
our observations here are expected to have interesting applications beyond $4d$ $\mathcal{N}=1$ theories corresponding to toric Calabi-Yau 3-folds.

In fact, the two $4d$ $\mathcal{N}=1$ theories and brane tilings presented here 
belong to an infinite family of distinct brane tilings that share the same mesonic moduli space 
and the same quiver, all related by the `tilting' mutation.
Given the appearance of brane tilings in recent studies of dimer integrable systems \cite{2011arXiv1107.5588G, Eager:2011dp, Kho:2025fmp, Kho:2025jxk}, 
$2d$ $(0,2)$ supersymmetric gauge theories associated with toric Calabi-Yau 4-folds \cite{Franco:2015tna, Franco:2015tya, Franco:2016qxh, Franco:2022gvl, Kho:2023dcm, Franco:2023tyf, Carcamo:2025shw}, 
and birational transformations of toric Calabi-Yau 3- and 4-folds \cite{Ghim:2024asj, Ghim:2025zhs}, 
we expect the correspondence identified here to contribute to a broader understanding of the algebro-geometric structures encoded by brane tilings --  
not only as individual $4d$ $\mathcal{N}=1$ theories, but also as part of a wider landscape of supersymmetric gauge theories. 
We leave a systematic exploration of further examples and related directions for future work.
\\

\subsection*{Acknowledgements}

The authors would like to thank D. Ghim, A. Hanany, V. Jejjala, J. Kwon, N. Lee and B. Suzzoni for discussions and comments.
S.-J. L. is supported by IBS-R003-D1. He also would like to thank UNIST for hospitality during important stages of this work.
R.-K. S. is supported by an Outstanding Young Scientist Grant (RS-2025-00516583) of the National Research Foundation of Korea (NRF).
He is also partly supported by the BK21 Program (``Next Generation Education Program for Mathematical Sciences'', 4299990414089) funded by the Ministry of Education in Korea and the National Research Foundation of Korea (NRF).
He is grateful to the Simons Center for Geometry and Physics at Stony Brook University, 
the Korea Institute for Advanced Study, 
the Institute of Mathematics of Academia Sinica, 
the Asia Pacific Center for Theoretical Physics, 
the University of M\"unster, the Kavli Institute for Theoretical Physics at UC Santa Barbara,
and the Center for Geometry and Physics at the Institute of Basic Science for hospitality during various stages of this work.

\bibliographystyle{jhep}
\bibliography{mybib}

\providecommand{\href}[2]{#2}\begingroup\raggedright\begin{thebibliography}{10}

\bibitem{Franco:2005rj}
S.~Franco, A.~Hanany, K.~D. Kennaway, D.~Vegh and B.~Wecht, \emph{{Brane dimers
  and quiver gauge theories}},
  \href{http://dx.doi.org/10.1088/1126-6708/2006/01/096}{\emph{JHEP} {\bf 01}
  (2006) 096}, [\href{http://arxiv.org/abs/hep-th/0504110}{{\tt
  hep-th/0504110}}].

\bibitem{Hanany:2005ve}
A.~Hanany and K.~D. Kennaway, \emph{{Dimer models and toric diagrams}},
  \href{http://arxiv.org/abs/hep-th/0503149}{{\tt hep-th/0503149}}.

\bibitem{Franco:2005sm}
S.~Franco, A.~Hanany, D.~Martelli, J.~Sparks, D.~Vegh and B.~Wecht,
  \emph{{Gauge theories from toric geometry and brane tilings}},
  \href{http://dx.doi.org/10.1088/1126-6708/2006/01/128}{\emph{JHEP} {\bf 01}
  (2006) 128}, [\href{http://arxiv.org/abs/hep-th/0505211}{{\tt
  hep-th/0505211}}].

\bibitem{2003math.....10326K}
R.~{Kenyon}, \emph{{An introduction to the dimer model}},
  \href{http://dx.doi.org/10.48550/arXiv.math/0310326}{\emph{arXiv Mathematics
  e-prints} (Oct., 2003) math/0310326},
  [\href{http://arxiv.org/abs/math/0310326}{{\tt math/0310326}}].

\bibitem{kasteleyn1967graph}
P.~Kasteleyn, \emph{Graph theory and crystal physics}, {\emph{Graph theory and
  theoretical physics} (1967) 43--110}.

\bibitem{Seiberg:1994pq}
N.~Seiberg, \emph{{Electric - magnetic duality in supersymmetric nonAbelian
  gauge theories}},
  \href{http://dx.doi.org/10.1016/0550-3213(94)00023-8}{\emph{Nucl. Phys. B}
  {\bf 435} (1995) 129--146}, [\href{http://arxiv.org/abs/hep-th/9411149}{{\tt
  hep-th/9411149}}].

\bibitem{Feng:2000mi}
B.~Feng, A.~Hanany and Y.-H. He, \emph{{D-brane gauge theories from toric
  singularities and toric duality}},
  \href{http://dx.doi.org/10.1016/S0550-3213(00)00699-4}{\emph{Nucl. Phys. B}
  {\bf 595} (2001) 165--200}, [\href{http://arxiv.org/abs/hep-th/0003085}{{\tt
  hep-th/0003085}}].

\bibitem{Feng:2001bn}
B.~Feng, A.~Hanany, Y.-H. He and A.~M. Uranga, \emph{{Toric duality as Seiberg
  duality and brane diamonds}},
  \href{http://dx.doi.org/10.1088/1126-6708/2001/12/035}{\emph{JHEP} {\bf 12}
  (2001) 035}, [\href{http://arxiv.org/abs/hep-th/0109063}{{\tt
  hep-th/0109063}}].

\bibitem{Feng:2001xr}
B.~Feng, A.~Hanany and Y.-H. He, \emph{{Phase structure of D-brane gauge
  theories and toric duality}},
  \href{http://dx.doi.org/10.1088/1126-6708/2001/08/040}{\emph{JHEP} {\bf 08}
  (2001) 040}, [\href{http://arxiv.org/abs/hep-th/0104259}{{\tt
  hep-th/0104259}}].

\bibitem{Feng:2002zw}
B.~Feng, S.~Franco, A.~Hanany and Y.-H. He, \emph{{Symmetries of toric
  duality}}, \href{http://dx.doi.org/10.1088/1126-6708/2002/12/076}{\emph{JHEP}
  {\bf 12} (2002) 076}, [\href{http://arxiv.org/abs/hep-th/0205144}{{\tt
  hep-th/0205144}}].

\bibitem{2011arXiv1107.5588G}
A.~B. {Goncharov} and R.~{Kenyon}, \emph{{Dimers and cluster integrable
  systems}}, \href{http://dx.doi.org/10.48550/arXiv.1107.5588}{\emph{arXiv
  e-prints} (July, 2011) arXiv:1107.5588},
  [\href{http://arxiv.org/abs/1107.5588}{{\tt 1107.5588}}].

\bibitem{CIUCU199834}
M.~Ciucu, \emph{A complementation theorem for perfect matchings of graphs
  having a cellular completion},
  \href{http://dx.doi.org/https://doi.org/10.1006/jcta.1997.2799}{\emph{Journal
  of Combinatorial Theory, Series A} {\bf 81} (1998) 34--68}.

\bibitem{1999math......3025K}
R.~W. {Kenyon}, J.~G. {Propp} and D.~B. {Wilson}, \emph{{Trees and Matchings}},
  \href{http://dx.doi.org/10.48550/arXiv.math/9903025}{\emph{arXiv Mathematics
  e-prints} (Mar., 1999) math/9903025},
  [\href{http://arxiv.org/abs/math/9903025}{{\tt math/9903025}}].

\bibitem{Butti:2007jv}
A.~Butti, D.~Forcella, A.~Hanany, D.~Vegh and A.~Zaffaroni, \emph{{Counting
  Chiral Operators in Quiver Gauge Theories}},
  \href{http://dx.doi.org/10.1088/1126-6708/2007/11/092}{\emph{JHEP} {\bf 11}
  (2007) 092}, [\href{http://arxiv.org/abs/0705.2771}{{\tt 0705.2771}}].

\bibitem{Forcella:2008bb}
D.~Forcella, A.~Hanany, Y.-H. He and A.~Zaffaroni, \emph{{The Master Space of
  N=1 Gauge Theories}},
  \href{http://dx.doi.org/10.1088/1126-6708/2008/08/012}{\emph{JHEP} {\bf 08}
  (2008) 012}, [\href{http://arxiv.org/abs/0801.1585}{{\tt 0801.1585}}].

\bibitem{Forcella:2008eh}
D.~Forcella, A.~Hanany, Y.-H. He and A.~Zaffaroni, \emph{{Mastering the Master
  Space}}, \href{http://dx.doi.org/10.1007/s11005-008-0255-6}{\emph{Lett. Math.
  Phys.} {\bf 85} (2008) 163--171}, [\href{http://arxiv.org/abs/0801.3477}{{\tt
  0801.3477}}].

\bibitem{Bianchi:2014qma}
M.~Bianchi, S.~Cremonesi, A.~Hanany, J.~F. Morales, D.~Ricci~Pacifici and R.-K.
  Seong, \emph{{Mass-deformed Brane Tilings}},
  \href{http://dx.doi.org/10.1007/JHEP10(2014)027}{\emph{JHEP} {\bf 10} (2014)
  027}, [\href{http://arxiv.org/abs/1408.1957}{{\tt 1408.1957}}].

\bibitem{2022SIGMA..18..030H}
A.~{Higashitani} and Y.~{Nakajima}, \emph{{Deformations of Dimer Models}},
  \href{http://dx.doi.org/10.3842/SIGMA.2022.030}{\emph{SIGMA} {\bf 18} (Apr.,
  2022) 030}, [\href{http://arxiv.org/abs/1903.01636}{{\tt 1903.01636}}].

\bibitem{Franco:2023flw}
S.~Franco and R.-K. Seong, \emph{{Twin theories, polytope mutations and quivers
  for GTPs}}, \href{http://dx.doi.org/10.1007/JHEP07(2023)034}{\emph{JHEP} {\bf
  07} (2023) 034}, [\href{http://arxiv.org/abs/2302.10951}{{\tt 2302.10951}}].

\bibitem{Cremonesi:2023psg}
S.~Cremonesi and J.~S{\'a}, \emph{{Zig-zag deformations of toric quiver gauge
  theories. Part I. Reflexive polytopes}},
  \href{http://dx.doi.org/10.1007/JHEP05(2024)114}{\emph{JHEP} {\bf 05} (2024)
  114}, [\href{http://arxiv.org/abs/2312.13909}{{\tt 2312.13909}}].

\bibitem{Arias-Tamargo:2024fjt}
G.~Arias-Tamargo, S.~Franco and D.~Rodr{\'\i}guez-G{\'o}mez, \emph{{The
  geometry of GTPs and 5d SCFTs}},
  \href{http://dx.doi.org/10.1007/JHEP07(2024)159}{\emph{JHEP} {\bf 07} (2024)
  159}, [\href{http://arxiv.org/abs/2403.09776}{{\tt 2403.09776}}].

\bibitem{Franco:2023mkw}
S.~Franco and D.~Rodriguez-Gomez, \emph{{Quiver tails and brane webs}},
  \href{http://dx.doi.org/10.1007/JHEP10(2024)118}{\emph{JHEP} {\bf 10} (2024)
  118}, [\href{http://arxiv.org/abs/2310.10724}{{\tt 2310.10724}}].

\bibitem{CarrenoBolla:2024fxy}
I.~Carre{\~n}o~Bolla, S.~Franco and D.~Rodr{\'\i}guez-G{\'o}mez, \emph{{The 5d
  tangram: brane webs, 7-branes and primitive T-cones}},
  \href{http://dx.doi.org/10.1007/JHEP05(2025)175}{\emph{JHEP} {\bf 05} (2025)
  175}, [\href{http://arxiv.org/abs/2411.01510}{{\tt 2411.01510}}].

\bibitem{Kho:2025fmp}
M.~Kho, N.~Lee and R.-K. Seong, \emph{{Birational transformations on dimer
  integrable systems}}, \href{http://dx.doi.org/10.1103/1mll-95cs}{\emph{Phys.
  Rev. D} {\bf 112} (2025) L041901},
  [\href{http://arxiv.org/abs/2504.21081}{{\tt 2504.21081}}].

\bibitem{Hanany:2012vc}
A.~Hanany and R.-K. Seong, \emph{{Brane Tilings and Specular Duality}},
  \href{http://dx.doi.org/10.1007/JHEP08(2012)107}{\emph{JHEP} {\bf 08} (2012)
  107}, [\href{http://arxiv.org/abs/1206.2386}{{\tt 1206.2386}}].

\bibitem{1997AIHPB..33..591K}
R.~{Kenyon}, \emph{{Local statistics of lattice dimers}},
  \href{http://dx.doi.org/10.1016/S0246-0203(97)80106-9}{\emph{Annales de
  L'Institut Henri Poincare Section (B) Probability and Statistics} {\bf 33}
  (Jan., 1997) 591--618}, [\href{http://arxiv.org/abs/math/0105054}{{\tt
  math/0105054}}].

\bibitem{fulton1993introduction}
W.~Fulton, \emph{Introduction to toric varieties}.
\newblock No.~131. Princeton university press, 1993.

\bibitem{cox2011graduate}
D.~Cox, J.~Little and H.~Schenck, \emph{Graduate studies in mathematics},
  2011.

\bibitem{Witten:1993yc}
E.~Witten, \emph{{Phases of N=2 theories in two-dimensions}},
  \href{http://dx.doi.org/10.1016/0550-3213(93)90033-L}{\emph{Nucl. Phys. B}
  {\bf 403} (1993) 159--222}, [\href{http://arxiv.org/abs/hep-th/9301042}{{\tt
  hep-th/9301042}}].

\bibitem{Benvenuti:2006qr}
S.~Benvenuti, B.~Feng, A.~Hanany and Y.-H. He, \emph{{Counting BPS Operators in
  Gauge Theories: Quivers, Syzygies and Plethystics}},
  \href{http://dx.doi.org/10.1088/1126-6708/2007/11/050}{\emph{JHEP} {\bf 11}
  (2007) 050}, [\href{http://arxiv.org/abs/hep-th/0608050}{{\tt
  hep-th/0608050}}].

\bibitem{Hanany:2006uc}
A.~Hanany and C.~Romelsberger, \emph{{Counting BPS operators in the chiral ring
  of N=2 supersymmetric gauge theories or N=2 braine surgery}},
  \href{http://dx.doi.org/10.4310/ATMP.2007.v11.n6.a4}{\emph{Adv. Theor. Math.
  Phys.} {\bf 11} (2007) 1091--1112},
  [\href{http://arxiv.org/abs/hep-th/0611346}{{\tt hep-th/0611346}}].

\bibitem{Feng:2007ur}
B.~Feng, A.~Hanany and Y.-H. He, \emph{{Counting gauge invariants: The
  Plethystic program}},
  \href{http://dx.doi.org/10.1088/1126-6708/2007/03/090}{\emph{JHEP} {\bf 03}
  (2007) 090}, [\href{http://arxiv.org/abs/hep-th/0701063}{{\tt
  hep-th/0701063}}].

\bibitem{Intriligator:2003jj}
K.~A. Intriligator and B.~Wecht, \emph{{The Exact superconformal R symmetry
  maximizes a}},
  \href{http://dx.doi.org/10.1016/S0550-3213(03)00459-0}{\emph{Nucl. Phys. B}
  {\bf 667} (2003) 183--200}, [\href{http://arxiv.org/abs/hep-th/0304128}{{\tt
  hep-th/0304128}}].

\bibitem{Butti:2005vn}
A.~Butti and A.~Zaffaroni, \emph{{R-charges from toric diagrams and the
  equivalence of a-maximization and Z-minimization}},
  \href{http://dx.doi.org/10.1088/1126-6708/2005/11/019}{\emph{JHEP} {\bf 11}
  (2005) 019}, [\href{http://arxiv.org/abs/hep-th/0506232}{{\tt
  hep-th/0506232}}].

\bibitem{Butti:2005ps}
A.~Butti and A.~Zaffaroni, \emph{{From toric geometry to quiver gauge theory:
  The Equivalence of a-maximization and Z-minimization}},
  \href{http://dx.doi.org/10.1002/prop.200510276}{\emph{Fortsch. Phys.} {\bf
  54} (2006) 309--316}, [\href{http://arxiv.org/abs/hep-th/0512240}{{\tt
  hep-th/0512240}}].

\bibitem{stanley1978hilbert}
R.~P. Stanley, \emph{Hilbert functions of graded algebras}, {\emph{Advances in
  Mathematics} {\bf 28} (1978) 57--83}.

\bibitem{Martelli:2006yb}
D.~Martelli, J.~Sparks and S.-T. Yau, \emph{{Sasaki-Einstein manifolds and
  volume minimisation}},
  \href{http://dx.doi.org/10.1007/s00220-008-0479-4}{\emph{Commun. Math. Phys.}
  {\bf 280} (2008) 611--673}, [\href{http://arxiv.org/abs/hep-th/0603021}{{\tt
  hep-th/0603021}}].

\bibitem{Martelli:2005tp}
D.~Martelli, J.~Sparks and S.-T. Yau, \emph{{The Geometric dual of
  a-maximisation for Toric Sasaki-Einstein manifolds}},
  \href{http://dx.doi.org/10.1007/s00220-006-0087-0}{\emph{Commun. Math. Phys.}
  {\bf 268} (2006) 39--65}, [\href{http://arxiv.org/abs/hep-th/0503183}{{\tt
  hep-th/0503183}}].

\bibitem{Butti:2006au}
A.~Butti, D.~Forcella and A.~Zaffaroni, \emph{{Counting BPS baryonic operators
  in CFTs with Sasaki-Einstein duals}},
  \href{http://dx.doi.org/10.1088/1126-6708/2007/06/069}{\emph{JHEP} {\bf 06}
  (2007) 069}, [\href{http://arxiv.org/abs/hep-th/0611229}{{\tt
  hep-th/0611229}}].

\bibitem{Bao:2024nyu}
J.~Bao, E.~Choi, Y.-H. He, R.-K. Seong and S.-T. Yau, \emph{{Futaki invariants
  and reflexive polygons}}, \href{http://dx.doi.org/10.1063/5.0263001}{\emph{J.
  Math. Phys.} {\bf 66} (2025) 102302},
  [\href{http://arxiv.org/abs/2410.18476}{{\tt 2410.18476}}].

\bibitem{Hanany:2005ss}
A.~Hanany and D.~Vegh, \emph{{Quivers, tilings, branes and rhombi}},
  \href{http://dx.doi.org/10.1088/1126-6708/2007/10/029}{\emph{JHEP} {\bf 10}
  (2007) 029}, [\href{http://arxiv.org/abs/hep-th/0511063}{{\tt
  hep-th/0511063}}].

\bibitem{2009arXiv0901.4662B}
N.~{Broomhead}, \emph{{Dimer models and Calabi-Yau algebras}},
  \href{http://dx.doi.org/10.48550/arXiv.0901.4662}{\emph{arXiv e-prints}
  (Jan., 2009) arXiv:0901.4662}, [\href{http://arxiv.org/abs/0901.4662}{{\tt
  0901.4662}}].

\bibitem{Closset:2019juk}
C.~Closset and M.~Del~Zotto, \emph{{On 5D SCFTs and their BPS quivers. Part I:
  B-branes and brane tilings}},
  \href{http://dx.doi.org/10.4310/ATMP.2022.v26.n1.a2}{\emph{Adv. Theor. Math.
  Phys.} {\bf 26} (2022) 37--142}, [\href{http://arxiv.org/abs/1912.13502}{{\tt
  1912.13502}}].

\bibitem{Eager:2011dp}
R.~Eager, S.~Franco and K.~Schaeffer, \emph{{Dimer Models and Integrable
  Systems}}, \href{http://dx.doi.org/10.1007/JHEP06(2012)106}{\emph{JHEP} {\bf
  06} (2012) 106}, [\href{http://arxiv.org/abs/1107.1244}{{\tt 1107.1244}}].

\bibitem{Kho:2025jxk}
M.~Kho, N.~Lee and R.-K. Seong, \emph{{Classification and Birational
  Equivalence of Dimer Integrable Systems for Reflexive Polygons}},
  \href{http://arxiv.org/abs/2510.12290}{{\tt 2510.12290}}.

\bibitem{Franco:2015tna}
S.~Franco, D.~Ghim, S.~Lee, R.-K. Seong and D.~Yokoyama, \emph{{2d (0,2) Quiver
  Gauge Theories and D-Branes}},
  \href{http://dx.doi.org/10.1007/JHEP09(2015)072}{\emph{JHEP} {\bf 09} (2015)
  072}, [\href{http://arxiv.org/abs/1506.03818}{{\tt 1506.03818}}].

\bibitem{Franco:2015tya}
S.~Franco, S.~Lee and R.-K. Seong, \emph{{Brane Brick Models, Toric Calabi-Yau
  4-Folds and 2d (0,2) Quivers}},
  \href{http://dx.doi.org/10.1007/JHEP02(2016)047}{\emph{JHEP} {\bf 02} (2016)
  047}, [\href{http://arxiv.org/abs/1510.01744}{{\tt 1510.01744}}].

\bibitem{Franco:2016qxh}
S.~Franco, S.~Lee, R.-K. Seong and C.~Vafa, \emph{{Brane Brick Models in the
  Mirror}}, \href{http://dx.doi.org/10.1007/JHEP02(2017)106}{\emph{JHEP} {\bf
  02} (2017) 106}, [\href{http://arxiv.org/abs/1609.01723}{{\tt 1609.01723}}].

\bibitem{Franco:2022gvl}
S.~Franco and R.-K. Seong, \emph{{Fano 3-folds, reflexive polytopes and brane
  brick models}}, \href{http://dx.doi.org/10.1007/JHEP08(2022)008}{\emph{JHEP}
  {\bf 08} (2022) 008}, [\href{http://arxiv.org/abs/2203.15816}{{\tt
  2203.15816}}].

\bibitem{Kho:2023dcm}
M.~Kho and R.-K. Seong, \emph{{On the master space for brane brick models}},
  \href{http://dx.doi.org/10.1007/JHEP09(2023)150}{\emph{JHEP} {\bf 09} (2023)
  150}, [\href{http://arxiv.org/abs/2306.16616}{{\tt 2306.16616}}].

\bibitem{Franco:2023tyf}
S.~Franco, D.~Ghim, G.~P. Goulas and R.-K. Seong, \emph{{Mass deformations of
  brane brick models}},
  \href{http://dx.doi.org/10.1007/JHEP09(2023)176}{\emph{JHEP} {\bf 09} (2023)
  176}, [\href{http://arxiv.org/abs/2307.03220}{{\tt 2307.03220}}].

\bibitem{Carcamo:2025shw}
M.~Carcamo, S.~Franco, D.~Ghim, G.~P. Goulas and R.-K. Seong, \emph{{Relevant
  Deformations, Brane Brick Models and Triality}},
  \href{http://arxiv.org/abs/2510.05517}{{\tt 2510.05517}}.

\bibitem{Ghim:2024asj}
D.~Ghim, M.~Kho and R.-K. Seong, \emph{{Combinatorial and algebraic mutations
  of toric Fano 3-folds and mass deformations of 2d(0,2) quiver gauge
  theories}}, \href{http://dx.doi.org/10.1103/PhysRevD.110.086001}{\emph{Phys.
  Rev. D} {\bf 110} (2024) 086001},
  [\href{http://arxiv.org/abs/2407.19924}{{\tt 2407.19924}}].

\bibitem{Ghim:2025zhs}
D.~Ghim, M.~Kho and R.-K. Seong, \emph{{Birational transformations and 2d (0,
  2) quiver gauge theories beyond toric Fano 3-folds}},
  \href{http://dx.doi.org/10.1007/JHEP06(2025)032}{\emph{JHEP} {\bf 06} (2025)
  032}, [\href{http://arxiv.org/abs/2502.08741}{{\tt 2502.08741}}].

\end{thebibliography}\endgroup

\end{document}